\documentclass[useAMS,usenatbib,usegraphicx]{mn2e}

\title[Transverse Proximity Effect]{The transverse proximity effect in
the $z \sim 2$ Lyman-$\alpha$ forest suggests
QSO episodic lifetimes of $\sim 1$~Myr}

\author[Kirkman \etal]{David Kirkman\thanks{E-mail: dkirkman@ucsd.edu}
  and David Tytler
  \\
  Center for Astrophysics and Space Sciences,
  University of California San Diego,
  La Jolla, CA, 92093-0424
}

\usepackage{color}
\usepackage{graphicx}
 
\newcommand{\lya}{\mbox{Ly$\alpha$}}
\newcommand{\lyb}{\mbox{Ly$\beta$}}

\newcommand{\kms}{\mbox{km s$^{-1}$}}

\newcommand{\msun}{$M_{\odot}~$}

\newcommand{\etal}{{\it et al.}}
\newcommand{\lyaf} {\lya\ forest}

\newif\ifdraftmodep
\draftmodepfalse

\newif\ifapjp
\apjpfalse

\newcommand{\dfig}[2]{
\begin{figure}
\includegraphics[width=80mm]{#1}
\caption{#2}
\end{figure}}

\newcommand{\dfigl}[2]{
\begin{figure}
\includegraphics[angle=270, width=80mm]{#1}
\caption{#2}
\end{figure}}

\newcommand{\dfigls}[3]{
\begin{figure}
\mbox{
  \includegraphics[angle=270, width=80mm]{#1}
}
\mbox {
  \includegraphics[angle=270, width=80mm]{#2}
}
\caption{#3}
\end{figure}}

\newcommand{\cc}[1]{\multicolumn{1}{c}{#1}}

\begin{document}

\date{\today}

\maketitle

\begin{abstract}
  We look for signs of the H~I transverse proximity effect in the spectra of
  130 QSO pairs, most with transverse separations in the plane of the
  sky of 0.1 -- 3 Mpc at $z \sim 2.2$.  We expected to see a decrease
  in \lyaf\ H~I absorption in the spectrum of background QSOs near the
  position of foreground QSOs. Instead we see no change in the
  absorption in front of the foreground QSOs, and we see evidence for
  a 50\% increase in the absorption out to 6 Mpc behind the foreground
  QSOs.  Further, we see no change in the H~I absorption along the
  line-of-sight to the foreground QSOs, the normal line-of-sight
  proximity effect.  We may account for the lack of change in the H~I
  absorption if the effect of extra UV photons is cancelled by higher
  gas density around QSOs.  If so, the increase in absorption behind
  the QSOs then suggests that the higher gas density there is not
  cancelled by the UV radiation from the QSOs. We can explain our observations
  if QSOs have had their current UV luminosities for less than
  approximately a million years, a time scale that has been suggested
  for accretion disk instabilities and gas depletion.
\end{abstract}


\section{Introduction}

Quasars are the most luminous known objects in the universe at 1 Ryd,
at least among objects that are luminous more than a few hours.
Consequently, they are expected to have a profound effect on neutral
hydrogen (H~I) in the nearby intergalactic medium (IGM).  This is
because in most of the volume of the IGM the H~I is expected to be
optically thin and in photoionization equilibrium with the
metagalactic UV background (UVB), so the enormous UV flux from the QSO
will significantly alter the photoionization equilibrium and reduce
the amount of neutral hydrogen.  The UV flux from a typical QSO at
$z=2$ should dominate the UVB as far away as 5 Mpc from the QSO.  The
resulting decrease in the optical depth of the \lyaf\ at redshifts
near to the systemic redshift of the QSO is known as the proximity
effect.

We define two proximity effects: the line-of-sight proximity effect
where the H~I optical depth is measured in the spectrum of the QSO
that is the source of the ionizing radiation, and the transverse
proximity effect where the H~I optical depth is measured in the
spectrum of an object which is near to the QSO in the plane of the
sky and further away.  Some other groups call the transverse
proximity effect the foreground proximity effect.  The line-of-sight
proximity effect is expected to be present in the spectra of all QSOs
unless there is an intervening optically thick absorber or the
episodic lifetime of the QSO is shorter than the inverse H~I
ionization rate in the IGM (0.1 Myr), i.e. the time required for each
H~I atom to experience one photoionization.

We use ``episodic lifetime'' in the same sense as
\citet{tp:martini2004a}, meaning the lifetime of the current outburst
and not the integrated ``on'' time of the QSO.  If QSOs have many
episodes of high UV luminosity, separated by off states with low
luminosity, the total QSO lifetime may be much longer than the
duration of the current UV luminous episode \citep{tp:martini2004a}.
An episodic lifetime less than a few Myr will significantly change the
appearance of the transverse proximity effect
(\citealt[Fig. 3]{tp:adelberger2004a}, \citealt{tp:schirber2004a},
\citealt[Fig. 1]{tp:visbal2008a}, and \citealt[Fig. 24]{tp:tytler2008a}).

The purpose of this paper is to measure the transverse proximity
effect in a large sample of QSO pairs that are separated in the plane
of the sky by $< 3$ Mpc.  We also measure the line-of-sight proximity
effect and compare it to the transverse effect because this helps us
explore the environments of the QSOs, anisotropic emission and episodic
lifetimes.

The line-of-sight proximity effect has been detected numerous times in
the spectra of tens of the brightest QSOs known at redshifts 1.6 -- 4.
It was first seen by \citet{carswell82} and \citet{tp:tytler1987a} in
the spectra of 7 and 19 QSOs, mostly at $z \sim 2$. They noticed a
decrease in the number of \lya\ lines at redshifts similar to QSO
emission redshifts.  Like most early papers, they used a sample of
\lya\ lines with rest frame equivalent widths exceeding a fixed
minimum (0.32~\AA ) and they excluded lines at redshifts that show
metal lines in moderate resolution spectra.  The line-of-sight
proximity effect can be seen in high resolution spectra of individual
QSOs at high redshift \citep{carswell87, giallongo96} where there is
more \lya\ absorption.  If QSOs are located in typical regions of the
IGM, then the amount of H~I absorption that is missing and the
distance from QSOs at which this occurs depends only upon the ratio of
the flux of ionizing photons from the QSO to the UVB.  Hence
\citet{bajtlik88} were able to use the effect to estimate the
intensity of the UVB in the Lyman continuum near the Lyman Limit, and
others have improved upon this method \citep{tp:scott2000a}.

Recently, \citet{tp:guimaraes2007a} used ESI spectra to examine the
proximity effect of 45 $z > 4$ QSOs.  They found significantly more
absorption than expected based upon the luminosity of the QSOs: they
conclude that the QSOs must reside in regions were the IGM is
over-dense by a factor of $\sim 5$.  A similar result was previously
reported by \citet{tp:rollinde2005a}.

We expect the gas near to QSOs to have higher than the average density
in the IGM because QSOs are in galaxies and the halo masses are large.
\citet{croom05a} estimated QSO halo masses of $4.2 \pm 2.3 \times
10^{12}$ solar masses in the 2QZ sample at all redshifts.
\citet{coil07b} also found a mean mass of $\sim 3 \times 10^{12}$ at
$0.7 < z < 1.4$.  Less directly, \citet{kim07a} use the distribution
of H~I absorption seen in background QSOs to estimate the masses of
foreground QSO halos.  They find a mean mass of $\log M =
12.48^{+0.53}_{-0.89}$ in solar units for QSOs at $z=3$ with an absolute
G-band magnitude $-27.5$, a factor of 20 above the mass of LBGs.
We also expect the gas near QSOs to have higher density
than the average IGM because \citet[Fig 14]{tp:adelberger2004a}
sees excess \lya\ absorption within 1 Mpc proper of LBGs.

The idea that QSOs are found in relatively dense environments is also
supported by the discovery that absorption is often produced when a
sightline passes within 100~kpc of a QSO. Absorbers are more often seen
in this case than when we look directly at a QSO. \citet{bowen06a}
discovered this effect with Mg~II absorption in four out of four QSOs,
while \citet{tp:hennawi2006a} saw the same for Lyman limit systems
(LLS) and Damped \lya\ lines (DLAs), and we have also seen the effect
in metal line systems with a super-set of the data examined here
\citep{tp:tytler2008a}.  \citet{hennawi07} conclude that QSOs live in
dense environments, and that the UV flux from the QSO
photo-evaporates LLS along the line-of-sight.  But because of either
anisotropic emission or short QSO episodic lifetimes, some of the LLS
in the transverse direction are spared or are less likely to be
photo-evaporated.  \citet{tp:wild2008a} estimate that QSOs destroy
nearby absorbers to comoving distances of 0.3 Mpc for C~IV systems,
and 0.8 Mpc for Mg~II systems.

For ions other than Hydrogen, enhanced ionization near to a foreground
QSO has been reported by several authors.  \citet{tp:jakobsen2003a}
find a significant lack of HeII absorption in the spectrum of
Q0302-003 at the redshift of a foreground QSO located $\sim 3$ Mpc
from the Q0302-003 line-of-sight.  Similarly, \citet{tp:worseck2007a}
examined both the HeII absorption and the H~I absorption towards HE
2347-4342, and found evidence for a hard ionizing spectrum near the
redshifts of 14 low luminosity foreground QSOs.
\citet{tp:goncalves2007a} have also detected unusual high ionization
absorption systems (e.g. O~VI) indicative of a transverse proximity
effect in high resolution spectra of the QSO triplet KP76, KP77, and
KP78.

In contrast, an H~I transverse proximity effect has yet to be
detected, with perhaps one exception \citep{tp:gallerani2008}.  It is
interesting in part because \citet{tp:adelberger2004a} has pointed out
that the transverse proximity effect is one of the best ways to
explore changes in the UV luminosity over several Myr.  Both
\citet{tp:wang2005a} and \citet{tp:visbal2008a} discuss how we may
obtain similar information from afterglows or light-echoes from QSOs
that were previously luminous.

\citet{tp:liske2001a} detected the line-of-sight proximity effect in a
group of 10 QSOs near $z = 2.9$ with transverse separations of
10--40\arcmin.  But with the exception of the sightline that passes about
10\arcmin\ from four separate foreground QSOs, they do not detect a
transverse proximity effect.  They conclude that QSOs emit at least a
factor of 1.4 less in the plane of the sky than they do along the
line-of-sight.  \citet{tp:schirber2004a} also did not detect the
transverse proximity effect, using three pairs of SDSS spectra
separated by $\sim$ 2 -- 4\arcmin.

\citet{tp:croft2004a} analyzed a sample of 325 QSOs with SDSS
spectra. He also failed to detect a transverse proximity, although his
sample was less sensitive than ours because it had almost no QSO pairs
as close as ours.  \citet{tp:croft2004a} actually observed an increase
in the mean absorption near to QSOs in the transverse direction, at a
level much higher than expected by his simulations which placed QSOs
in high density regions.  This may be related to the \citet{bowen06a},
\citet{tp:hennawi2006a} and \citet{tp:tytler2008a} results, since it
is likely that the \citet{tp:croft2004a} simulations, like most
simulations, underestimate the number of LLS and DLA systems
\citep{tp:tytler2008b}.  Recently \citet{tp:gallerani2008} have
reported an increased density of flux spikes near a
foreground QSO at z=5.6 that they model as
the first example of the transverse H~I proximity effect.

In this paper we look for the H~I transverse proximity effect in a
large sample of sightlines passing within 0.1 -- 3 Mpc of QSOs.  We do
not see the obvious transverse proximity effect expected if the UV
flux from QSOs is long lived and emitted isotropically.  But we do
detect asymmetry in the absorption around the QSOs, which may be a
result of a combination of enhanced IGM density within a few Mpc of
QSOs, combined with short episodic lifetimes for QSO outbursts.  We
can understand our observations if the QSOs were much less UV luminous
$\sim$ 1 Myr ago than they are today.  For all of the calculations in
this paper we take $H_0 = 71$ \kms\ Mpc$^{-1}$, $\Omega_m = 0.27$, and
$\Omega_\Lambda = 0.73$.  All distances in this paper are proper,
unless noted otherwise.

\section{Data}
\label{dataprep}

\begin{table*}
  \caption{\label{obstab} QSO Pairs.
    For each member of the pair we give the RA and Dec (J2000), the
    adopted emission redshift $z_{\rm em}$, magnitude, and the SNR 
    per \AA\ (S) at 1260 \AA\ in the QSO rest frame.  The background QSO 
    is listed first.  $\theta$ is the angular separation on the sky between 
    the 2 QSOs in arcseconds, and $b$ is the impact parameter in
    proper Mpc between the two sightlines at the $z_{\rm em}$ of the foreground
    QSO.  $L$ is the estimated Lyman Limit luminosity in units of $10^{30}$ ergs
    s$^{-1}$ Hz$^{-1}$.  $\omega_{\rm max}$ is the expected ratio of the 
    UV flux from the foreground QSO to the UVB, at the point of closest 
    approach of the background sightline to the foreground QSO.  
  }

\begin{tabular}{rrllr  rrllr rrrr}
\cc{RA} & \cc{Dec} & \cc{$z_{\rm em}$} & \cc{mag} & \cc{S} &
\cc{RA} & \cc{Dec} & \cc{$z_{\rm em}$} & \cc{mag} & \cc{S} &
\cc{$\theta$} & \cc{$b$} & \cc{$L$} & \cc{$\omega_{\rm max}$}  \\
\hline
   11 16 10.7 & $+$41 18 14.5 &  2.995 & 19.5  g &   9 &    11 16 11.7 & $+$41 18 21.5 &  2.983 & 18.6  g &  19 &  13.7 &  0.10 &        13 & 2272.5 \cr
   09 09 23.1 & $+$00 02 03.9 &  1.893 & 20.2  g &  36 &    09 09 24.0 & $+$00 02 11.0 &  1.878 & 16.8  g & 102 &  15.0 &  0.12 &        22 & 2715.3 \cr
   03 13 25.5 & $-$31 41 54.3 &  2.075 & 20.1 bJ &  33 &    03 13 24.4 & $-$31 41 44.9 &  1.965           & 20.0 bJ &  31 &  17.0 &  0.14 &       1.3 & 123.1 \cr
   02 18 21.4 & $-$29 53 40.9 &  2.067 & 20.1 bJ &  66 &    02 18 23.0 & $-$29 53 31.3 &  1.919           & 19.2 bJ &  77 &  22.0 &  0.18 &       2.6 & 149.7 \cr
   11 07 27.1 & $+$00 34 07.3 &  1.890 & 20.2  g &  29 &    11 07 25.7 & $+$00 33 53.6 &  1.881 & 19.1  g &  58 &  24.8 &  0.21 &       2.7 & 119.5 \cr
   21 48 36.6 & $-$29 40 54.2 &  2.098 & 20.1 bJ & 102 &    21 48 34.9 & $-$29 41 09.9 &  1.822           & 19.6 bJ &  36 &  26.7 &  0.22 &       1.6 &  62.4 \cr
   13 06 34.2 & $+$29 24 43.1 &  1.978 & 19.1  V &  30 &    13 06 35.4 & $+$29 25 05.9 &  1.938           & 20.6  V &  34 &  27.8 &  0.23 &      0.71 &  25.1 \cr
   21 43 07.0 & $-$44 50 47.6 &  3.270 & 21.3  V &  25 &    21 43 04.1 & $-$44 50 36.0 &  3.067           & 21.1  V &  24 &  33.2 &  0.25 &       1.1 &  33.0 \cr
   14 35 06.4 & $+$00 09 01.5 &  2.389 & 20.0 bJ &   4 &    14 35 08.3 & $+$00 08 44.4 &  2.384           & 20.1 bJ &   2 &  33.2 &  0.27 &         2 &  53.8 \cr
   23 53 13.0 & $-$27 26 09.4 &  2.308 & 20.1 bJ &  10 &    23 53 10.0 & $-$27 26 14.1 &  1.963           & 18.7 bJ &  52 &  40.3 &  0.33 &       4.4 &  74.1 \cr
   23 59 45.5 & $-$00 58 19.6 &  1.820 & 18.7  g &  77 &    23 59 44.1 & $-$00 57 38.2 &  1.795 & 19.5  g &  48 &  46.2 &  0.38 &       1.7 &  21.2 \cr
   23 09 11.9 & $-$27 32 27.1 &  1.932 & 19.4 bJ &  30 &    23 09 15.3 & $-$27 32 45.3 &  1.927           & 20.2 bJ &   4 &  49.5 &  0.41 &       1.1 &  12.3 \cr
   22 32 20.3 & $-$28 38 58.7 &  2.207 & 19.9 bJ &   7 &    22 32 23.4 & $-$28 38 29.9 &  2.073           & 20.3 bJ &  36 &  50.8 &  0.42 &       1.1 &  12.2 \cr
   03 06 43.7 & $-$30 11 07.5 &  2.127 & 19.8 bJ &  18 &    03 06 40.9 & $-$30 10 31.9 &  2.099           & 19.3 bJ &  35 &  51.2 &  0.42 &       2.9 &  30.6 \cr
   02 56 42.6 & $-$33 15 21.0 &  1.915 & 17.0  V &  35 &    02 56 47.0 & $-$33 15 27.0 &  1.872           & 16.5  V &  19 &  55.8 &  0.46 &        29 & 251.2 \cr
   09 27 47.3 & $+$29 07 20.7 &  2.304 & 18.6  g &  13 &    09 27 43.0 & $+$29 07 34.7 &  2.254 & 19.1  g &   9 &  57.4 &  0.47 &       3.8 &  33.2 \cr
   03 10 06.1 & $-$19 21 24.9 &  2.152 & 18.6  V &  74 &    03 10 09.0 & $-$19 22 08.1 &  2.129           & 19.1  V &  37 &  60.3 &  0.49 &       3.3 &  25.3 \cr
   21 42 25.9 & $-$44 20 17.0 &  3.242 & 18.7  V &  12 &    21 42 22.2 & $-$44 19 30.0 &  3.227           & 21.2  V &  59 &  61.5 &  0.46 &       1.1 &  10.0 \cr
   22 39 51.8 & $-$29 48 37.0 &  2.130 & 18.8 bJ &  81 &    22 39 48.6 & $-$29 47 48.7 &  2.071           & 20.2 bJ &  32 &  63.6 &  0.52 &       1.2 &   8.6 \cr
   08 15 18.3 & $+$06 06 04.3 &  2.536 & 20.2  g &   7 &    08 15 14.3 & $+$06 05 42.5 &  2.505 & 20.4  g &   4 &  64.1 &  0.51 &       1.6 &  11.5 \cr
   09 14 10.3 & $+$46 10 50.0 &  2.358 & 20.2  V &   5 &    09 14 04.1 & $+$46 10 44.9 &  2.191           & 21.0  V &   2 &  64.6 &  0.53 &      0.57 &   3.9 \cr
   10 41 29.3 & $+$56 30 23.0 &  2.267 & 19.0  g &  11 &    10 41 21.9 & $+$56 30 01.0 &  2.051 & 18.3  g &  16 &  65.1 &  0.54 &       6.7 &  44.3 \cr
   23 01 17.6 & $-$31 43 59.2 &  2.140 & 19.1 bJ &  65 &    23 01 12.4 & $-$31 43 45.0 &  1.991           & 19.7 bJ &  36 &  67.8 &  0.56 &       1.8 &  10.8 \cr
   10 16 05.8 & $+$40 40 05.8 &  2.995 & 20.5  g &   4 &    10 16 01.5 & $+$40 40 52.9 &  2.984 & 19.6  g &   8 &  68.2 &  0.52 &       5.3 &  36.7 \cr
   03 10 36.5 & $-$30 51 08.4 &  2.566 & 20.4 bJ &  13 &    03 10 41.1 & $-$30 50 27.5 &  2.546           & 19.5 bJ &  22 &  71.9 &  0.57 &       4.1 &  23.8 \cr
   14 57 56.3 & $+$57 44 46.9 &  2.125 & 19.7  g &   7 &    14 57 47.5 & $+$57 44 23.5 &  2.014 & 19.3  g &   9 &  73.6 &  0.61 &       2.6 &  13.2 \cr
   12 12 51.1 & $-$00 53 42.2 &  2.480 & 20.4 bJ &   1 &    12 12 56.1 & $-$00 53 36.5 &  2.467           & 20.8 bJ &   2 &  74.0 &  0.59 &       1.2 &   6.4 \cr
   00 08 52.7 & $-$29 00 44.1 &  2.656 & 19.1 bJ & 133 &    00 08 57.7 & $-$29 01 26.9 &  2.615           & 19.8 bJ &  89 &  78.5 &  0.62 &       3.4 &  16.5 \cr
   00 45 27.5 & $-$32 01 35.4 &  2.006 & 19.8 bJ &  45 &    00 45 26.5 & $-$32 00 16.9 &  1.896           & 19.0 bJ &  68 &  79.6 &  0.66 &       3.2 &  13.8 \cr
   01 02 57.4 & $-$27 53 38.8 &  1.820 & 19.3 bJ &  36 &    01 02 51.9 & $-$27 53 03.3 &  1.801           & 18.2  V &  51 &  81.1 &  0.67 &       5.5 &  23.0 \cr
   00 44 34.1 & $+$00 19 03.5 &  1.878 & 19.2  g &  21 &    00 44 39.3 & $+$00 18 22.8 &  1.875 & 18.3  g &  30 &  88.6 &  0.73 &       5.6 &  19.4 \cr
   15 45 34.6 & $+$51 12 28.0 &  2.458 & 19.5  g &   4 &    15 45 44.2 & $+$51 13 07.0 &  2.252 & 19.3  g &   6 &  98.3 &  0.80 &       3.2 &   9.4 \cr
   01 24 53.1 & $-$28 52 51.5 &  2.100 & 18.8 bJ &  78 &    01 24 56.4 & $-$28 51 21.0 &  1.998           & 19.0 bJ &  54 & 100.7 &  0.83 &       3.3 &   9.0 \cr
   15 19 19.4 & $+$23 46 02.0 &  1.907 & 16.4  V &  54 &    15 19 13.3 & $+$23 46 58.7 &  1.846           & 18.4  V &  26 & 101.3 &  0.84 &       4.8 &  12.7 \cr
   17 36 35.5 & $+$55 28 29.4 &  1.997 & 19.9  g &  39 &    17 36 26.7 & $+$55 27 20.7 &  1.831 & 20.4  g &  29 & 101.5 &  0.84 &      0.76 &   2.0 \cr
   11 06 11.1 & $+$13 56 00.0 &  3.912 & 21.5  g &   6 &    11 06 16.7 & $+$13 54 58.6 &  3.854 & 21.3  g &   6 & 101.7 &  0.71 &       3.5 &  13.0 \cr
   13 39 39.0 & $+$00 10 22.0 &  2.124 & 19.3  g &  57 &    13 39 45.4 & $+$00 09 45.0 &  1.879 & 19.3  g &  36 & 102.9 &  0.85 &       2.2 &   5.8 \cr
   16 23 24.8 & $+$33 10 49.8 &  2.593 & 18.1  g &  21 &    16 23 23.7 & $+$33 12 32.6 &  2.420 & 19.0  g &  13 & 103.7 &  0.83 &         5 &  13.6 \cr
   17 19 32.9 & $+$29 19 29.0 &  3.303 & 20.3  g &   5 &    17 19 37.9 & $+$29 18 05.0 &  3.075 & 20.7  g &   3 & 106.5 &  0.81 &       2.1 &   6.0 \cr
   14 22 39.9 & $+$42 02 20.4 &  3.228 & 19.5  g &   9 &    14 22 49.2 & $+$42 02 46.2 &  3.077 & 20.1  g &   4 & 106.9 &  0.81 &       3.7 &  10.6 \cr
   13 21 47.7 & $+$01 06 04.8 &  2.138 & 20.1  g &  54 &    13 21 54.3 & $+$01 06 51.9 &  1.983           & 20.1  g &  22 & 110.6 &  0.91 &       1.2 &   2.7 \cr
   17 27 56.4 & $+$58 21 55.7 &  2.373 & 19.3  g &  49 &    17 28 06.8 & $+$58 20 39.2 &  2.019 & 19.4  g &  37 & 111.6 &  0.92 &       2.4 &   5.3 \cr
   11 08 19.1 & $-$00 58 24.0 &  4.567 & 23.5  g &   7 &    11 08 13.9 & $-$00 59 44.5 &  4.027 & 21.0  g &  11 & 113.0 &  0.78 &       5.3 &  16.6 \cr
   11 52 00.5 & $+$45 17 41.4 &  2.406 & 19.2  g &  10 &    11 52 10.4 & $+$45 18 25.8 &  2.311 & 19.1  g &   9 & 113.4 &  0.92 &         4 &   8.9 \cr
   16 12 45.6 & $+$23 58 00.0 &  2.040 & 19.6  g &  22 &    16 12 37.9 & $+$23 57 09.0 &  2.018           & 19.4  V &  19 & 117.2 &  0.97 &       2.3 &   4.7 \cr
   15 09 32.2 & $+$50 57 51.5 &  2.377 & 18.8  g &  11 &    15 09 25.6 & $+$50 56 09.3 &  2.375 & 19.0  g &  12 & 119.7 &  0.96 &       4.7 &   9.4 \cr
   13 46 21.4 & $-$00 38 05.0 &  1.895 & 20.1 bJ &  12 &    13 46 25.6 & $-$00 39 47.0 &  1.851           & 19.3 bJ &  10 & 119.9 &  0.99 &       2.2 &   4.1 \cr
   02 09 54.8 & $-$10 02 23.0 &  1.980 & 19.6  r &  16 &    02 10 00.1 & $-$10 03 54.0 &  1.979 & 19.3  g &  51 & 120.0 &  0.99 &       2.5 &   4.8 \cr
   15 08 38.1 & $+$60 35 40.1 &  2.188 & 19.0  g &  13 &    15 08 27.7 & $+$60 34 07.4 &  1.901 & 18.1  g &  20 & 120.5 &  1.00 &       6.9 &  13.0 \cr
   10 05 41.3 & $+$57 05 44.0 &  2.324 & 18.1  g &  12 &    10 05 38.5 & $+$57 07 44.0 &  1.874 & 18.9  g &   4 & 122.1 &  1.01 &       3.2 &   5.9 \cr
   00 55 57.5 & $-$32 55 39.0 &  2.257 & 19.6  V &  41 &    00 56 05.3 & $-$32 56 51.1 &  2.134           & 19.6  V &   5 & 122.5 &  1.00 &       2.1 &   3.9 \cr
   10 19 22.9 & $+$55 24 31.0 &  3.728 & 21.5  g &   3 &    10 19 37.0 & $+$55 23 55.0 &  3.236 & 20.5  g &   4 & 125.4 &  0.93 &         3 &   6.5 \cr
   15 48 50.2 & $+$53 38 43.0 &  2.192 & 19.8  g &   4 &    15 48 40.8 & $+$53 37 08.6 &  2.172 & 18.8  g &  10 & 126.1 &  1.03 &       4.8 &   8.4 \cr
   11 04 11.6 & $+$02 46 55.0 &  2.534 & 18.3  g &  20 &    11 04 03.0 & $+$02 47 20.0 &  2.368 & 21.0  g &   2 & 131.3 &  1.06 &      0.73 &   1.2 \cr
   20 53 02.9 & $-$01 02 25.0 &  3.217 & 21.1  g &   3 &    20 53 03.7 & $-$01 04 42.0 &  3.118 & 20.6  g &   4 & 137.5 &  1.04 &       2.4 &   4.3 \cr
   12 13 10.7 & $+$12 07 15.1 &  3.487 & 20.3  g &   4 &    12 13 03.3 & $+$12 08 39.0 &  3.377 & 20.8  g &   5 & 137.9 &  1.01 &       3.1 &   5.8 \cr
   07 55 35.6 & $+$40 58 02.9 &  2.423 & 19.0  g &  17 &    07 55 45.6 & $+$40 56 43.6 &  2.346 & 19.3  g &  12 & 138.2 &  1.11 &       3.4 &   5.2 \cr
   11 43 17.0 & $+$13 24 00.8 &  2.520 & 18.9  g &  12 &    11 43 23.4 & $+$13 25 42.0 &  2.517 & 18.8  g &  13 & 138.3 &  1.10 &         7 &  10.9 \cr
   23 26 14.3 & $-$29 37 22.3 &  2.393 & 19.1 bJ &  25 &    23 26 03.5 & $-$29 37 40.4 &  2.318           & 20.6 bJ &  20 & 141.2 &  1.14 &       1.2 &   1.7 \cr
   13 02 16.9 & $-$03 38 03.7 &  3.758 & 21.1  g &   6 &    13 02 08.2 & $-$03 37 10.5 &  3.718 & 20.6  g &   6 & 141.2 &  1.00 &         6 &  11.3 \cr

\end{tabular}
\end{table*}

\setcounter{table}{0}
\begin{table*}
\caption{{\it continued}}
\begin{tabular}{rrllr  rrllr rrrr}
\cc{RA} & \cc{Dec} & \cc{$z_{\rm em}$} & \cc{mag} & \cc{S} &
\cc{RA} & \cc{Dec} & \cc{$z_{\rm em}$} & \cc{mag} & \cc{S} &
\cc{$\theta$} & \cc{$b$} & \cc{$L$} & \cc{$\omega_{\rm max}$}  \\
\hline
   15 59 22.7 & $+$52 00 27.0 &  3.122 & 20.3  g &   5 &    15 59 17.4 & $+$52 02 44.0 &  3.035 & 19.2  g &  12 & 145.5 &  1.10 &       8.1 &  12.5 \cr
   03 33 24.8 & $-$06 10 03.4 &  2.157 & 18.8  g &  38 &    03 33 20.9 & $-$06 12 16.8 &  2.042 & 18.4  g &  49 & 145.7 &  1.20 &       6.1 &   8.0 \cr
   13 12 13.3 & $+$00 02 31.2 &  2.892 & 20.9  g &   2 &    13 12 13.8 & $+$00 00 03.0 &  2.675 & 19.5  g &   9 & 148.4 &  1.17 &       4.3 &   6.0 \cr
   11 45 53.7 & $-$00 33 04.5 &  2.055 & 20.1  g &  16 &    11 45 47.5 & $-$00 31 06.7 &  2.048 & 18.7  g &  48 & 149.3 &  1.23 &       4.6 &   5.8 \cr
   09 16 03.4 & $+$33 09 31.8 &  3.148 & 20.3  g &   8 &    09 16 11.0 & $+$33 11 30.5 &  3.103 & 19.5  g &  12 & 152.5 &  1.15 &       6.6 &   9.3 \cr
   23 31 32.8 & $+$01 06 20.9 &  2.644 & 18.9  g &  11 &    23 31 39.7 & $+$01 04 27.0 &  2.239 & 18.5  g &  13 & 154.0 &  1.25 &       6.6 &   7.9 \cr
   13 24 11.6 & $+$03 20 50.0 &  3.670 & 22.2  g &   1 &    13 24 01.5 & $+$03 20 20.0 &  3.371 & 20.0  g &   3 & 154.2 &  1.13 &       6.5 &   9.5 \cr
   17 18 37.2 & $+$30 28 52.0 &  2.049 & 19.7  g &   5 &    17 18 45.0 & $+$30 26 47.0 &  2.037 & 19.9  g &   4 & 160.6 &  1.32 &       1.5 &   1.6 \cr
   13 54 38.4 & $+$59 31 34.0 &  3.000 & 20.4  g &   4 &    13 54 42.9 & $+$59 28 56.0 &  2.560 & 19.9  g &   5 & 161.7 &  1.28 &       2.7 &   3.1 \cr
   11 19 31.1 & $+$60 49 21.0 &  2.648 & 18.3  g &  24 &    11 19 28.9 & $+$60 46 37.0 &  2.305 & 19.0  g &   8 & 164.8 &  1.33 &       4.4 &   4.6 \cr
   14 33 56.3 & $+$23 22 22.8 &  4.145 & 22.5  g &   5 &    14 34 08.3 & $+$23 22 30.0 &  4.010 & 21.7  g &   7 & 166.1 &  1.14 &       2.7 &   4.0 \cr
   16 50 51.1 & $+$34 43 10.0 &  2.007 & 18.5  g &  21 &    16 50 43.3 & $+$34 45 30.0 &  1.987 & 20.3  g &   2 & 169.8 &  1.40 &         1 &   1.0 \cr
   08 52 37.9 & $+$26 37 58.6 &  3.317 & 19.8  g &   9 &    08 52 32.2 & $+$26 35 26.2 &  3.217 & 20.8  g &   4 & 170.9 &  1.27 &       2.2 &   2.6 \cr
   13 37 57.9 & $+$02 18 20.9 &  3.322 & 18.6  g &  19 &    13 37 56.3 & $+$02 15 30.1 &  2.319 & 19.2  g &  11 & 172.3 &  1.39 &       3.7 &   3.6 \cr
   16 25 57.7 & $+$26 44 43.4 &  2.611 & 19.1  g &  11 &    16 25 48.7 & $+$26 46 58.6 &  2.539           & 17.3  g &  22 & 180.3 &  1.43 &        29 &  26.2 \cr
   01 06 57.9 & $-$08 55 00.1 &  2.364 & 18.2  g &  15 &    01 06 58.4 & $-$08 58 01.9 &  1.829 & 18.5  g &  10 & 181.9 &  1.51 &       4.4 &   3.6 \cr
   09 44 53.8 & $+$50 43 00.0 &  3.768 & 20.6  g &   8 &    09 45 08.0 & $+$50 40 57.0 &  3.748 & 19.8  g &  12 & 182.6 &  1.29 &        13 &  14.5 \cr
   14 29 33.0 & $+$63 14 12.4 &  2.754 & 20.2  g &   7 &    14 29 51.9 & $+$63 16 31.9 &  2.409 & 18.7  g &  19 & 188.9 &  1.52 &       6.5 &   5.3 \cr
   10 40 19.1 & $+$32 21 56.4 &  2.654 & 20.6  g &   5 &    10 40 04.0 & $+$32 21 50.6 &  2.633 & 19.0  g &  12 & 191.0 &  1.51 &       6.6 &   5.5 \cr
   08 03 05.8 & $+$50 32 15.3 &  3.242 & 20.5  g &   5 &    08 03 21.3 & $+$50 34 17.4 &  3.239 & 20.3  g &   6 & 191.1 &  1.42 &       3.6 &   3.4 \cr
   02 48 40.1 & $-$28 03 32.4 &  2.228 & 19.6 bJ &   6 &    02 48 25.6 & $-$28 03 55.4 &  2.143           & 19.5 bJ &  48 & 193.8 &  1.59 &       2.6 &   2.0 \cr
   09 46 42.4 & $+$33 07 54.8 &  2.543 & 19.0  g &  13 &    09 46 56.2 & $+$33 06 25.8 &  2.538 & 19.0  g &  12 & 194.2 &  1.54 &         6 &   4.7 \cr
   12 09 10.7 & $+$11 35 45.2 &  3.122 & 19.3  g &  11 &    12 09 17.9 & $+$11 38 30.4 &  3.118 & 17.7  g &  29 & 196.4 &  1.48 &        35 &  30.3 \cr
   10 00 54.4 & $+$45 03 29.0 &  2.650 & 20.1  g &   5 &    10 00 52.2 & $+$45 00 11.0 &  2.570 & 20.2  g &   5 & 199.3 &  1.58 &       2.1 &   1.6 \cr
   14 30 06.4 & $-$01 20 20.0 &  3.249 & 20.6  g &   4 &    14 29 57.1 & $-$01 17 57.0 &  3.117 & 19.7  g &   7 & 199.7 &  1.50 &       5.4 &   4.5 \cr
   14 11 30.7 & $+$62 22 48.6 &  2.308 & 18.8  g &   8 &    14 11 08.0 & $+$62 24 52.2 &  2.264 & 21.2  g &   1 & 200.4 &  1.63 &      0.56 &   0.4 \cr
   14 22 09.7 & $+$46 59 32.5 &  3.809 & 21.5  g &   5 &    14 21 50.0 & $+$46 59 38.6 &  3.676 & 21.6  g &   2 & 201.9 &  1.44 &       2.3 &   2.1 \cr
   15 37 29.5 & $+$58 32 24.0 &  3.080 & 21.0  g &   3 &    15 37 15.7 & $+$58 29 33.0 &  2.581 & 17.8  g &  25 & 202.3 &  1.60 &        19 &  13.8 \cr
   08 33 26.8 & $+$08 15 52.0 &  2.574 & 18.0  g &  20 &    08 33 21.6 & $+$08 12 38.6 &  2.516 & 20.5  g &   4 & 208.3 &  1.66 &       1.5 &   1.0 \cr
   17 17 30.7 & $+$26 22 27.0 &  2.201 & 18.5  g &  18 &    17 17 15.2 & $+$26 21 48.0 &  1.943 & 18.6  g &   7 & 211.9 &  1.75 &       4.6 &   2.8 \cr
   03 40 23.5 & $+$00 31 11.8 &  1.913 & 20.1  g &   3 &    03 40 27.3 & $+$00 34 41.5 &  1.881 & 17.9  g &  18 & 217.3 &  1.80 &       8.1 &   4.7 \cr
   13 31 38.5 & $+$00 42 21.1 &  2.435 & 18.6  g &  13 &    13 31 25.9 & $+$00 44 14.0 &  2.030 & 18.9  g &   7 & 219.8 &  1.81 &       3.8 &   2.2 \cr
   11 09 27.2 & $+$55 41 20.0 &  3.465 & 19.7  g &   9 &    11 09 52.3 & $+$55 42 24.0 &  3.181 & 19.7  g &   8 & 221.6 &  1.66 &         6 &   4.1 \cr
   14 34 55.4 & $+$03 50 30.9 &  2.855 & 18.4  g &  20 &    14 35 00.3 & $+$03 54 03.5 &  2.490 & 19.6  g &   8 & 224.8 &  1.79 &       3.2 &   1.9 \cr
   08 07 35.0 & $+$23 51 26.4 &  3.773 & 21.7  g &   4 &    08 07 44.9 & $+$23 48 25.7 &  3.730 & 21.2  g &   4 & 225.9 &  1.60 &       3.5 &   2.6 \cr
   09 35 48.5 & $+$36 31 21.9 &  2.974 & 18.9  g &  14 &    09 35 31.8 & $+$36 33 17.6 &  2.867 & 18.5  g &  16 & 231.8 &  1.79 &        13 &   7.6 \cr
   12 38 31.5 & $+$44 32 58.2 &  3.323 & 20.2  g &   6 &    12 38 15.0 & $+$44 30 26.2 &  3.261 & 19.2  g &   8 & 232.2 &  1.72 &        11 &   6.8 \cr
   14 26 28.0 & $+$50 02 48.0 &  2.328 & 18.3  g &  20 &    14 26 05.8 & $+$50 04 26.0 &  2.244 & 17.9  g &  25 & 235.2 &  1.91 &        11 &   5.9 \cr
   13 47 55.7 & $+$00 39 35.0 &  3.816 & 20.5  g &   6 &    13 48 08.8 & $+$00 37 23.2 &  3.620 & 20.6  g &   8 & 236.7 &  1.70 &       5.6 &   3.6 \cr
   21 36 15.4 & $+$10 27 54.0 &  2.957 & 19.9  g &   8 &    21 36 29.4 & $+$10 29 52.0 &  2.554 & 18.2  g &  19 & 237.8 &  1.89 &        13 &   6.7 \cr
   10 54 16.5 & $+$51 27 24.6 &  2.372 & 18.8  g &   6 &    10 54 16.5 & $+$51 23 26.1 &  2.347 & 18.9  g &  14 & 238.5 &  1.92 &         5 &   2.5 \cr
   01 06 12.2 & $+$00 19 20.1 &  3.094 & 19.5  g &  13 &    01 06 16.1 & $+$00 15 24.0 &  3.029 & 20.7  g &   6 & 243.0 &  1.85 &         2 &   1.1 \cr
   16 43 30.1 & $+$30 55 41.8 &  2.734 & 19.4  g &   9 &    16 43 41.3 & $+$30 58 59.8 &  2.597 & 18.9  g &  10 & 244.5 &  1.93 &         7 &   3.5 \cr
   17 30 42.4 & $+$54 56 01.1 &  2.129 & 18.8  g &  34 &    17 30 14.7 & $+$54 56 57.5 &  2.117 & 20.1  g &  38 & 245.0 &  2.01 &       1.4 &   0.6 \cr
   14 19 19.5 & $+$57 45 13.0 &  3.341 & 19.9  g &   6 &    14 19 00.6 & $+$57 48 30.0 &  2.948 & 20.1  g &   5 & 248.3 &  1.90 &       3.2 &   1.7 \cr
   01 35 14.5 & $-$00 53 18.9 &  2.114 & 19.1  g &   8 &    01 35 21.0 & $-$00 57 18.2 &  2.085 & 19.1  g &   7 & 258.2 &  2.12 &       3.3 &   1.4 \cr
   10 38 49.3 & $+$55 13 37.5 &  3.858 & 21.4  g &   2 &    10 38 41.5 & $+$55 09 27.8 &  3.696 & 21.4  g &   3 & 258.5 &  1.84 &       2.8 &   1.6 \cr
   14 53 29.5 & $+$00 23 57.3 &  2.540 & 18.4  g &  20 &    14 53 38.0 & $+$00 20 10.5 &  1.858 & 19.3  g &   5 & 259.9 &  2.16 &       2.2 &   0.9 \cr
   08 25 40.1 & $+$35 44 14.0 &  3.853 & 20.1  g &  15 &    08 25 50.2 & $+$35 48 03.0 &  3.198 & 20.7  g &   4 & 259.9 &  1.94 &       2.4 &   1.2 \cr
   11 51 22.1 & $+$02 04 26.4 &  2.411 & 19.1  g &  13 &    11 51 38.0 & $+$02 06 10.4 &  2.264 & 20.2  g &   5 & 260.2 &  2.11 &       1.4 &   0.6 \cr
   12 19 33.3 & $+$00 32 26.4 &  2.886 & 19.6  g &  10 &    12 19 22.2 & $+$00 29 05.4 &  2.633 & 19.2  g &  13 & 260.7 &  2.05 &       5.5 &   2.4 \cr
   09 45 05.9 & $-$00 46 44.9 &  2.302 & 20.4  g &   6 &    09 44 54.2 & $-$00 43 30.4 &  2.298 & 19.2  g &  10 & 261.9 &  2.12 &       3.6 &   1.5 \cr
   08 04 00.3 & $+$30 20 46.0 &  3.451 & 21.0  g &   7 &    08 03 42.0 & $+$30 22 54.0 &  2.028 & 16.5  g &  53 & 269.2 &  2.22 &        35 &  13.2 \cr
   08 36 59.8 & $+$35 10 19.4 &  3.321 & 19.9  g &   7 &    08 37 00.8 & $+$35 05 50.2 &  3.311 & 18.7  g &  16 & 269.5 &  1.99 &        18 &   8.8 \cr
   00 59 51.7 & $-$08 44 23.8 &  2.158 & 18.5  g &  15 &    00 59 34.1 & $-$08 43 13.1 &  2.082 & 18.3  g &  16 & 269.9 &  2.22 &         7 &   2.7 \cr
   08 59 59.1 & $+$02 05 19.7 &  2.980 & 18.5  g &  19 &    08 59 56.8 & $+$02 09 52.8 &  2.235 & 20.2  g &   3 & 275.3 &  2.24 &       1.4 &   0.5 \cr
   11 26 34.3 & $-$01 24 36.0 &  3.750 & 20.1  g &   7 &    11 26 17.4 & $-$01 26 32.0 &  3.626 & 19.8  g &   8 & 278.7 &  2.00 &        12 &   5.5 \cr
   11 11 14.1 & $+$01 20 34.4 &  2.168 & 18.5  g &  14 &    11 11 31.3 & $+$01 22 25.0 &  2.007 & 19.1  g &   4 & 280.5 &  2.31 &       3.1 &   1.1 \cr
   15 33 48.3 & $+$50 31 28.0 &  2.225 & 18.5  g &   5 &    15 34 12.7 & $+$50 34 05.0 &  2.126 & 17.6  g &  14 & 280.6 &  2.30 &        14 &   4.9 \cr
   23 37 56.6 & $-$10 20 00.1 &  2.442 & 19.1  g &  10 &    23 38 15.4 & $-$10 19 17.2 &  2.279 & 21.0  g &   1 & 281.8 &  2.28 &      0.68 &   0.2 \cr
   08 54 06.1 & $+$42 38 10.0 &  2.394 & 18.6  g &  14 &    08 54 15.4 & $+$42 42 34.0 &  2.176 & 19.3  g &  10 & 283.2 &  2.31 &         3 &   1.1 \cr
   14 16 47.6 & $+$63 02 51.0 &  2.035 & 18.1  g &  14 &    14 16 50.8 & $+$63 07 35.0 &  1.965 & 19.1  g &   4 & 284.8 &  2.35 &       2.9 &   1.0 \cr
   22 47 40.2 & $-$09 15 11.8 &  4.167 & 22.8  g &   1 &    22 47 21.1 & $-$09 15 48.7 &  4.130 & 22.6  g &   1 & 285.3 &  1.94 &       1.3 &   0.7 \cr
   15 00 23.5 & $+$61 47 29.0 &  3.004 & 20.0  g &   5 &    15 00 58.7 & $+$61 45 06.0 &  2.593 & 21.0  g &   3 & 287.8 &  2.28 &         1 &   0.4 \cr

\end{tabular}
\end{table*}

\setcounter{table}{0}
\begin{table*}
\caption{{\it continued}}
\begin{tabular}{rrllr  rrllr rrrr}
\cc{RA} & \cc{Dec} & \cc{$z_{\rm em}$} & \cc{mag} & \cc{S} &
\cc{RA} & \cc{Dec} & \cc{$z_{\rm em}$} & \cc{mag} & \cc{S} &
\cc{$\theta$} & \cc{$b$} & \cc{$L$} & \cc{$\omega_{\rm max}$}  \\
\hline
   08 30 53.0 & $+$38 12 43.0 &  3.165 & 20.5  g &   5 &    08 31 15.9 & $+$38 14 24.0 &  3.069 & 20.7  g &   5 & 288.1 &  2.18 &       2.1 &   0.8 \cr
   21 36 19.4 & $+$00 41 31.0 &  2.032 & 18.4  g &  55 &    21 36 38.6 & $+$00 41 54.0 &  1.949 & 17.1  g &  65 & 288.9 &  2.39 &        18 &   6.0 \cr
   10 42 53.4 & $-$00 13 00.9 &  2.957 & 18.9  g &  22 &    10 42 43.1 & $-$00 17 06.0 &  1.977 & 18.3  g &  12 & 289.8 &  2.39 &       6.2 &   2.1 \cr
   14 20 46.0 & $-$00 05 18.0 &  2.198 & 19.4  g &  58 &    14 20 55.6 & $-$00 09 40.0 &  2.194 & 19.5  g &  57 & 299.2 &  2.44 &       2.5 &   0.8 \cr
   17 30 30.2 & $+$60 19 47.4 &  2.219 & 19.0  g &   9 &    17 29 43.4 & $+$60 21 54.2 &  1.927 & 20.3  g &   2 & 370.0 &  3.06 &      0.94 &   0.2 \cr
   14 59 01.3 & $+$00 21 23.7 &  1.994 & 18.6  g &  18 &    14 58 38.0 & $+$00 24 18.0 &  1.896 & 18.9  g &   8 & 389.7 &  3.23 &       3.3 &   0.6 \cr

\end{tabular}
\end{table*}

Our data set consists of the \lyaf\ regions of the spectra of 130
close pairs of QSOs, which we list in Table \ref{obstab}.  We obtained
the spectra for a project to measure the Alcock-Paczynski effect with
the \lyaf.  The QSOs were selected from a list of all known QSOs with
NED magnitudes (the precise band varied, typically $g$, $B_J$, or $V$)
less than 22.  The pairs were then selected to have similar redshifts
to maximise the amount of overlapping \lyaf\ absorption.  The pairs we
chose to observe were selected based primarily on the estimated amount
of time it would take to get usable spectra of both members of the
pair, with a preference for observing pairs at close angular
separations.  We also added all pairs with usable \lyaf\ spectra in
the SDSS.  Except for nine of the pairs, at the time of the selection
there were no other known QSOs with $V < 22$ within 5\arcmin\ of either
member of any pair.  For the nine, there was one other QSO within 5\arcmin\
of one member the pair.

These data provide \lyaf\ spectra between redshifts $1.7 < z < 4.5$,
with a median redshift of $z_{\rm med} = 2.2$.  In all but two cases,
the angular separation between the two sightlines is $\theta < 5
\arcmin$, the median separation is $\theta_{\rm med} = 154$\arcsec,
and the linear separation is $b < 3$ Mpc.

The spectra were obtained from either Keck+LRIS (76 spectra),
Lick+KAST (26 spectra), NOAO 4m telescopes (10 spectra), or from the
SDSS DR5 archive (146 spectra).  The LRIS and KAST spectra were taken
with a narrow slit in a variety of conditions, so they do not
provide absolute spectrophotometry.  The typical SNR per \AA\ of our
spectra is 11.2 at a rest wavelength of 1260 \AA.  The resolution of
the LRIS spectra varies between 83 \kms\ and 234 \kms, the KAST spectra
have a resolution of 250 \kms, and the SDSS spectra have a resolution
of 165 \kms.  In \citet{tp:tytler2008a} we give the instrumental
setups, wavelength regions, exposure times and an indication of the
SNR for the best spectrum for each QSO. We also list the metal lines
we find and their redshifts.

To investigate the foreground QSOs' proximity effect, we would ideally
like to isolate the pixels which are dominated by absorption with a
low to moderate optical depth from H~I in the IGM.  The absorption in
the \lyaf\ region of a spectrum can be described by three components
\citep{tytler04b}: the H~I absorption from low density regions of the
IGM (about 80\% of the total at $z=2$), the H~I absorption from LLS
and DLAs ($\sim 10$\%), and metal absorption ($\sim 10$\%).  The QSO
radiation field we describe in Section \ref{radsec} has a well defined
effect on optically thin H~I, but its effect on the optically thick
LLS and DLA in a spectrum is less clear.  Most metal absorption near
the 1215 \AA\ in the QSO rest frame will come from systems with
redshifts much different than that of the QSO, and will hence be
unaffected by the QSO radiation field.  The notable exception is
Si~III (1206), which is often the strongest metal line in the \lyaf.
There will be no Si~III (1206) absorption superimposed on the \lyaf\
within $\sim 2500$ \kms\ (about 12 Mpc) of the QSO where the radiation
field is expected to be strongest, but it will be superimposed on the
\lyaf\ absorption at further distances.

To avoid contamination from non-\lyaf\ absorption, we have attempted
to identify all of the Lyman limit and DLA systems in the spectra.  We
have also attempted to identify obvious metal lines in the \lyaf\
associated with the systems with high H~I columns.  We specifically
searched for Si~III (1206).  We flagged each pixel in the \lyaf\ which
was found to be influenced by either a large column density \lya\
line, or by a known metal line.  We also flagged all pixels that seem
to be affected by BAL outflows.  The pixels flagged by this procedure
were discarded and not used again in our analysis.

We used an interactive program, described in \citet{kirkman05a}, to
manually place a B-spline continuum on each QSO.  We can control the
position of the continuum by moving a small number of control points
(the B-spline knots).  The number of knots is not constrained -- we
can add them as required to get a good fit.  However, we have a strong
bias for a smooth continuum except near the positions of known
emission lines.  From past experience (e.g. \citet{kirkman05a},
\citet{tytler04c}) we expect that this procedure should produce good
results on our low resolution spectra for $z < 2.5$, but that we will
likely place the continuum level too low at higher redshifts as line
blanketing increases leaving fewer pixels near the unabsorbed
continuum level.  With higher resolution spectra our continuum placement
procedure can be used at higher redshifts, but with the spectra used
here our continuum levels are likely to have large systematic errors
at the higher redshifts.  In Figures \ref{contin1} and \ref{contin2},
we show the spectra and continuum for two of our QSO pairs.

\dfigl{spec1.ps}{\label{contin1} Spectra of 22 39 51.8
  $-$29 48 37 (top, Keck+LRIS 1200/3400 grism) and 22 39 48.6 $-$29 47
  49 (bottom, Keck+LRIS 400/3400 grism).  Our continuum fit is
  indicated as a smooth blue line.  The wavelength units are \AA, and 
  the flux is linear $f_\lambda$ from zero at the lowest major axis mark.
  }

\dfigl{spec2.ps}{\label{contin2} As Figure \ref{contin1} but this
  time showing spectra of 14 57 56.6 +57 44 46 and 14 57 47.6 +57 44
  23.5, both observed with the Lick+KAST 830/3880 grism}

\subsection{Systemic Redshifts}
\label{redshifts}

A proximity effect analysis is very sensitive to the adopted
redshift for each QSO whose environment is being probed.  We would
like to use the systemic redshifts of the galaxies that host the QSOs,
but we know that the peaks of the main UV emission lines give
redshifts that are systematically smaller than these systemic
redshifts by many hundreds of \kms , and in some cases over 1000~\kms,
with a large QSO-to-QSO scatter \citep{gaskell82,tytler92}.  A $1000$
\kms\ redshift error corresponds to a 5~Mpc position error at $z=2$,
which is large compared to the region where we expect the QSO UV radiation
to be larger than the UVB, which is 3~Mpc for our median luminosity foreground
QSO.

We have measured the positions of the peaks of up to three emission
lines (\lya, CIV and MgII) for most QSOs in an attempt to get better
estimates for the systemic redshifts.  We measure vacuum heliocentric
wavelengths, and we do not use a line when the line peak is obscured
by strong absorption or the peak is not in available spectra.

The velocity shifts that we find between different emission lines are
comparable but not the same as those found by others.  We find that
C~IV gives a redshift smaller than Mg~II by $ 753$ \kms, with a
standard deviation (QSO to QSO $\sigma$) of $676$ \kms from 27 QSOs.
\citet{tp:richards2002a} finds Mg~II$-$C~IV $= 824$ \kms, with $\sigma
= 511$ \kms\ from a subset of 3814 SDSS QSO spectra.  We find \lya\
gives a redshift smaller by $475$ ($\sigma = 455$ \kms) than Mg~II.
\citet{tp:guimaraes2007a} effectively adopt a C~IV/\lya\ offset of 750
\kms.  \citet{tp:dallaglio2008a} adopt systemic redshifts from the
weak SiII+O~I emission line assuming a rest wavelength of 1305.77 \AA.
For their eight spectra with C~IV emission line coverage, this
corresponds to C~IV at a redshift smaller by 1190 \kms.

We calculate our redshifts using rest wavelengths 1215.67 \AA\ for
\lya, 1549.06 for C~IV, and 2798.74 for Mg~II. If our spectra cover
Mg~II, we use the redshift from that line alone (28 of our foreground
QSOs). Otherwise, if our spectra cover C~IV, we assign a redshift for
C~IV alone and then increase the redshift by 753 \kms (78 QSOs).
Otherwise, we use the redshift from \lya\ increased by 475 \kms (20
QSOs).  For 4 QSOs we did not measure any of the emission lines and we
list and use the redshift listed in NED.  We also used this algorithm
to assign redshifts to the background QSOs.

\subsection{SDSS redshifts}
\label{sdssredshiftssec}

A large number of our foreground QSOs have SDSS spectra and the SDSS
project has given these QSOs redshifts derived using a template
spectrum in a manner that should give the systemic redshift. 
For the 41 QSOs where we have both Mg~II emission lines and SDSS
redshifts, our redshifts are larger by 574 \kms\ with a standard
deviation of 552 \kms.  We obtained the SDSS redshifts from the DR5
QSO catalogue \citep{tp:schneider2007a}.  

We only understand part of the difference in the redshifts.  Our
methods are considerably different.  The SDSS team fit a template
spectrum \citep{stoughton02} based on effective rest frame wavelengths
for emission lines from \citet{vandenberk01}.  The
\citet{vandenberk01} rest frame wavelengths are calculated assuming
that [O~III] gives the systemic redshift of the QSO.  About a third of
the difference is due to our differing rest frame wavelength for the
Mg~II emission line.  We assume that Mg~II is systemic, while
\citet{vandenberk01} assume [O~III] is systemic and then find
[O~III]$-$Mg~II$=-161$ \kms (Mg~II gives a higher redshift than
[O~III]).  We expect that the rest of the difference must come from
the differing methods, our using line peaks vs. SDSS fitting template
spectra.

We agree that [O~III] should better represent the systemic
redshift, but none of our QSOs have infrared spectra that can be used
to measure [O~III], and there is significant disagreement about the
[O~III] -- Mg~II offset.  \citet{tp:richards2002a}, also using SDSS
spectra, find [O~III]$-$Mg~II $=-97$ \kms, and \citet{tp:nestor2008a}
find [O~III] -- Mg~II $ = + 102$ \kms.  Unlike the SDSS spectra, which
are all at low redshift so that [O~III] is in the optical, the
\citet{tp:nestor2008a} were at $z \sim 2$ (like our QSOs) and have
[O~III] positions from NIR spectra taken by \citet{tp:mcintosh1999a}
and \citet{tp:scott2000a}.

\section{Method}

We look for the transverse proximity effect by measuring the mean
amount of absorption along the lines-of-sight to the background QSOs
at redshifts near to that of the foreground QSOs. We sum  the spectra of
all 130 QSOs to average over the random changes in the amount of absorption
in the IGM. The amount of absorption varies by factors of many over
short distances, as we move in and out of absorption lines, and there
are correlations that we have previously measured out to scales
of 150~Mpc \citep{tytler04c}.

We then compare the result to the expected amount of absorption for
this data set. We convert the QSO magnitudes into luminosities, and we
calculate the ionizing flux we expect from the foreground QSO at
various distances along each background sightline.  We end by
examining what the foreground QSOs do to the amount of absorption in
the line-of-sight to us.

Our methods differ from all early papers on the proximity effect
since they counted the number of \lya\ lines with rest frame
equivalent width exceeding some minimum, and they excluded \lya\ lines
at redshifts that showed metal lines. Hence individual strong lines
have a larger effect on our measurements than in the line counting
method.  We are also sensitive to the numerous weak lines that are
below the equivalent width threshold. Such lines are easier to see
where the SNR is higher, such as in the \lya\ emission line.  We will
be examining the changes in the relative amount of absorption as we
pass the foreground QSOs.  We are then insensitive to the total amount
of absorption per QSO and to global systematic errors in the continuum
level.  

We are sensitive to systematic errors caused by fitting the continuum
differently in different parts of a spectrum.  For example, the errors
in our continuum may be different in and far from emission lines.
This is not known to be a problem, but we can not rule out the
possibility.  Any systematic differences in continuum fitting over
emission lines will be most noticeable in our line-of-sight proximity
effect, which uses all of our data right through the \lya\ emission
line.  Any errors fitting the continuum to emission lines will be less
important for the transverse proximity effect. First, we restrict our
analysis to wavelengths $< 1200$~\AA\ far from the peak of the \lya\
line.  In \citet[Figs. 5 \& 6]{tytler04b} we showed that 1200~\AA\ is
far enough from emission line peaks to avoid unusually large continuum
level errors.  Second, for many of our pairs the foreground QSO
redshift corresponds to a wavelength well away from any emission lines
in the background QSO.

Several recent papers have detected the line-of-sight proximity effect
using optical depth instead of line counting: \citet{tp:liske2001a},
\citet{tp:schirber2004a}, \citet{tp:rollinde2005a},
\citet{tp:guimaraes2007a} and
\citet{tp:dallaglio2008a}. \citet{tp:guimaraes2007a} and
\citet{tp:liske2001a} used $\tau_{\rm eff}$, and their methods are similar
to the methods we use here. The other papers use statistics derived
from optical depth.  \citet{tp:croft2004a} utilized the mean flux in
an attempt to detect the transverse proximity effect, which is
directly equivalent to our method.

We quantify the amount of absorption in our spectra with DA = $1 -
F/C$, where $F$ is the flux and $C$ is the continuum level.
Equivalently, our DA values can be converted to effective optical
depth using = ${\rm DA} = 1 - e^{-\tau_{\rm eff}}$, where $\tau_{\rm
  eff}$ is an effective optical depth.  We calculate DA for each
individual pixel in a spectrum, excluding of course those pixels which
have been flagged for containing LLS, DLA, metal or BAL absorption.
Most of the figures in this paper show DA which has been binned in
some way, e.g. the DA for all the pixels in all the spectra within
some redshift interval.  In such cases, we find the mean DA for all
the pixels in that bin from one sightline, then we average the DA
values from the different sightlines.

The error we derive for a bin comes from the distribution of binned DA
values, with one DA value per sightline.  So when we resample, we
resample the sightlines, not the individual pixels.  We do the same
when we estimate the error on the mean from the dispersion of the DA
in individual sightlines.  We do this because if we went
pixel-by-pixel we would give much more weight to our higher resolution
spectra.  Also, taken sightline by sightline the binned DA values are
statistically independent, while taken pixel-by-pixel they are not,
because adjacent pixels in individual spectra are highly correlated.

DA evolves rapidly with redshift (see, for example
\citet{tp:kirkman2007a} and references therein).  In Figure
\ref{figzda} we show the redshift evolution of DA in our data.  We
observe significantly different absorption at $z > 2.5$ than we
measured from high resolution spectra in \citet{tp:kirkman2007a}.  We
are not surprised by this, because line blending in the \lyaf\
makes the continuum levels difficult to measure in low resolution
spectra at higher redshift.  In this case it seems that we have placed
the continua too low, giving too little DA, though the sample is also
getting small at high $z$ so some of the difference could be the
random fluctuations in the IGM \citep{tytler04c}.

\dfigl{zda.ps}{\label{figzda} The DA against redshift in this data
  set, including both the foreground and background sightlines.  The
  bins contain contributions from pixels between 1070 and 1170 \AA\ in
  the rest frame of each QSO.  The data histogram shows DA averaged
  over all QSOs in various redshift bins.  This is intended to
  approximate the DA from the low density IGM. We have masked out and
  ignore the \lya\ lines of LLS and DLAs and the metal absorption that
  we can identify.  The solid blue curve shows our fit to the DA
  measured in this sample (Equation \ref{daeq}).  The dashed curve
  shows the IGM DA measured in spectra intended for this purpose from
  \citet{tp:kirkman2007a}.
}

We fit the redshift evolution of DA in our data set with a simple
power series
\begin{equation}
\label{daeq}
 {\rm DA}(z) = -0.069 + 0.082 z + 0.006 z^2.
\end{equation}
To enable us to compare data at different redshifts, the final step in
our data preparation is to re-scale the DA values to those expected at $z=2$,
\begin{equation}
{\rm DA}_n = (1 - F/C) {\rm DA}(2.0)/ {\rm DA}(z),
\end{equation}
where DA$(z)$ is given by Equation \ref{daeq}.

\section{The expected line-of-sight and transverse proximity effects}
\label{radsec}

We define the photoionization enhancement $\omega$ at a particular
point in space due to the foreground QSO to be
\begin{equation}
\omega = {\Gamma_{\rm QSO} \over \Gamma_{\rm UVB}}
\end{equation}
where $\Gamma_{\rm QSO}$ is the photoionization rate (ionizations per
H~I atom per second) due to the
quasar radiation, and $\Gamma_{\rm UVB}$ is the photoionization rate
due to the UV background.  If other factors are equal,
the optical depth $\tau$ near the QSO is then given by
\begin{equation}
\label{taueq}
\tau = {\tau_0 \over (1 + \omega)}
\end{equation}
where $\tau_0$ is the optical depth that the IGM would have had in the
absence of the foreground QSO UV flux.  The linear scaling with
$1+\omega$ is formally only true for the actual optical depth. The
effective optical depth $\tau_{\rm eff}$ that is proportional to DA
may have a slightly different scaling.  For example, a high column
density line on the flat part of the curve of growth will have its
$\tau$ decreased when subjected to an enhanced radiation field, but
its equivalent width will remain unchanged.  However, we have shown in
\citet{jena05a} that at $z=2$ the effective optical depth also scales
like $\tau_{\rm eff} \propto \Gamma^{-1}$, so we substitute $\tau_{\rm
  eff}$ for $\tau $ in Equation \ref{taueq}.  We then calculate how
the absorption changes near the QSOs using DA $ = 1 - e^{-\tau_{\rm
    eff}}$.

Our assumption that $\tau_{\rm eff} = \tau_0 (1+ \omega )^{-1}$ will be
valid for unsaturated lines which make much of the absorption, but we
expect less sensitivity to $(1+ \omega )$ for other lines.  While
\citet{jena05a} found $\tau_{\rm eff} = \tau_0 (1+\omega )^{-1}$, the
spectra from their simulations with different $\Gamma $ were also
consistent with a range of relationships, including the $\tau_{\rm eff}
= \tau_0 (1+\omega )^{-0.69}$ from \citet{bolton05a}.  If we change the
index to $-0.7$, the expected reduction in DA near the QSOs is less,
by at most 0.016 which is the size of our errors on DA.

$\Gamma_{\rm QSO}$ at some distance $r = \sqrt{d^2 + b^2}$ from the QSO is given by
\begin{equation}
\label{gammaeq}
\Gamma_{\rm QSO} = \int^\infty_{\nu_0} {L_{\rm QSO}(\nu) \over 4 \pi r^2} 
   {\sigma_{\rm HI}(\nu) \over h \nu} d\nu
\end{equation}
where $L_{\rm QSO}(\nu)$ is the luminosity of the QSO as a function of
frequency, $\nu_0$ is the Lyman limit frequency, and the H~I
photoinization cross section \citep[Section 5.1]{tp:spitzer1978}
\begin{equation}
\sigma_{\rm HI}(\nu) = 6.3 \times 10^{-18} \left( {\nu \over \nu_0} \right)^{-2.75} {\rm cm}^2
\end{equation}
Both the coefficient in front and the exponent after the $\nu/\nu_0$
term include the approximate effects of the Gaunt factor near 1 Ryd.
Since we do not have direct observations of the Lyman limit regions
for most of our QSOs, we assume that the flux distribution of each QSO
is a power law $L_{\rm QSO}(\nu) = L_{\rm QSO} ( {\nu / \nu_0}
)^\alpha$, where $L_{\rm QSO}$ is the luminosity of the QSO at the
Lyman limit $\nu_0$.

Integrating Equation \ref{gammaeq} gives
\begin{equation}
\Gamma_{\rm QSO}^{12} = {0.951 \over 2.75 - \alpha}
{L_{\rm QSO} \over 4 \pi r^2} 10^{21} (10^{-12} {\rm s}^{-1}),
\end{equation}
For this equation we use $\alpha =
-0.5$ for the power law index to describe the continua of the QSOs.

We use $\Gamma_{\rm UVB} = 1.3 \times 10^{-12}$ s$^{-1}$ for all
redshifts.  This is the value we found in \citet{tp:tytler2008b} at
$z=2$ by matching the observed absorption in the \lyaf\ with the
absorption seen in large hydrodynamic simulations.  \citet{bolton05a}
found $\Gamma_{\rm UVB} = 1.3^{+0.8}_{-0.5} \times 10^{-12}$ s$^{-1}$
at $z=2$ using a similar method.  \citet{tp:scott2000a} use the
line-of-sight proximity effect to estimate $\Gamma_{\rm IGM} = 1.9 \pm
1 \times 10^{-12}$s$^{-1}$ using all absorption and some value between
0.9 and $1.9 \times 10^{-12}$s$^{-1}$ when they exclude absorption by
the \lya\ lines with associated metal lines.  \citet{haardt01a}
calculate $\Gamma_{\rm IGM} = 1.33 \pm 1 \times 10^{-12}$s$^{-1}$ when
they absorb the flux emitted by populations of QSOs and galaxies at
$z=1.9$.  The median redshift of our transverse proximity measurement
is $z = 2.2$ and the majority of our data is near that redshift. We
ignore the small changes we expect in $\Gamma_{\rm IGM}$ at other
redshifts.

\subsection{QSO Luminosities}
\label{lumsec}

We compute the luminosity of each QSO at the Lyman limit (in ergs
sec$^{-1}$ Hz$^{-1}$) from the observed flux density via the relation
\begin{equation}
L_{\rm QSO} = 4 \pi D_L^2(z) F_\nu / (1+z)
\end{equation}
where $F_\nu$ is the observed flux density at $\lambda = (1+z) 912$ \AA,
and $D_L(z)$ is the luminosity distance to redshift $z$,
\begin{equation}
D_L(z) = (1+z) {c \over H_0}
    \int_0^z {dz^\prime \over \sqrt{\Omega_m (1+z^\prime)^3 + \Omega_\Lambda}}
\end{equation}

We have estimated the Lyman continuum flux $F_\nu$ from a broadband
magnitude (either $g$, $B_J$, or $V$) for each QSO, since in general
we lack a direct measurement of $F_\nu$.  We estimate $F_\nu$ by
assuming each QSO has the spectrum of the HST composite QSO spectrum
\citep{zhang97}.  This is not ideal because there is significant
variability in the spectral slope of individual objects, but it should
be sufficient on average, which is what we need because we will always
be combining many sightlines.  To estimate $F_\nu$ for a given QSO, we
compute $m$, a synthetic AB magnitude for the redshifted composite spectrum
via the relation \citep{tp:fukugita96a}
\begin{equation}
\label{mageq}
m = -2.5 \log {\int d(\log \nu) A C_\nu(z) S_\nu \over 
               \int d(\log \nu) S_\nu} - 48.60
\end{equation}
where $C_\nu(z)$ is the HST composite QSO spectrum at a given
redshift, $S_\nu$ is the filter response, and $F_\nu = A C_\nu(z)$ is
the observed flux.  $A$ is a free parameter that is adjusted so that the 
synthetic magnitude is the same as the observed magnitude.  $C_\nu(z)$ 
is calculated by redshifting the composite QSO spectrum, and then removing 
flux to simulate the mean effects of \lyaf\ absorption, using the mean 
DA vs. redshift given in \citet{kirkman07a}.

Equation \ref{mageq} gives AB magnitudes, which are in the same system
as our $g$ band magnitudes.  $B_J$ and $V$ magnitudes are not on the
AB system, but the difference between AB and conventional magnitudes
for those filters is $< 0.15$ mag \citep{tp:fukugita96a}, and we
ignore the difference in this paper and assume that all magnitudes are
AB.  The situation is further complicated by the fact that our QSOs
with $B_J$ magnitudes are from the 2QZ, which has magnitudes
determined from APM scans of UKST photographic plates.  We do not have
a response curve for that combination of filter and emulsion.  Instead
we have used the filter-only response for the Tyson BJ filter on the
CTIO mosaic imager, which was designed to be consistent with the
photographic system.  For the V magnitudes we used the filter-only
response for the V filter on the CTIO mosaic.  For the $g$ magnitudes
we used the SDSS published filter+CCD+1.2 airmass atmosphere response.
The CCD and atmosphere change slowly through the filters, so their
primary effect is to suppress the entire response by a nearly constant
factor, and this has no net effect in Equation \ref{mageq}.

\subsection{Distances}
\label{distsec}

We compute the transverse distance $b$, the shortest distance in the plane of the sky
from the background sightline to the foreground QSO, with
\begin{equation}
b = {\phi \over (1+z)^2} D_L(z)
\end{equation}
where $\phi$ is the separation between the two QSOs in radians.

To calculate the ionizing flux expected from the foreground QSO at a
particular point in nearby space, we assume a Euclidean geometry and
calculate the distance between the QSO and the distance $r = \sqrt{b^2 + d^2}$,
where $d$ is the line-of-sight distance
\begin{equation}
d = {c \Delta z \over (1+z) H(z)},
\end{equation}
where $H(z)$ is the Hubble constant at the redshift of the foreground QSO, and
is given by
\begin{equation}
H(z) = H_0 \sqrt{\Omega_m (1+z)^3 + \Omega_\Lambda}.
\end{equation}
or 219.73 \kms\ at $z=2.2$.

\section{Observed Absorption near the Foreground QSOs}
\label{obsforsec}

In Figure \ref{abs00ned} we show the average absorption (expressed as
DA) near the foreground QSOs.  The top panel shows the absorption
observed in the line-of-sight towards the foreground QSOs, and the
lower panel shows the absorption observed in the line-of-sight towards
the background QSOs.  In both panels, the $x$-axis is the distance in
proper Mpc along a line-of-sight, with the origin at the
foreground QSOs, and negative distance behind the foreground QSOs.

The top panel of Figure \ref{abs00ned} contains contributions from
pixels with rest wavelength greater than 1070 \AA.  The DA value for
each pixel was scaled to $z=2$ using Equation \ref{daeq} and then
placed in the appropriate bin.  The bottom panel is computed in the
same way, except that only pixels with rest wavelengths between 1070
and 1200 \AA\ in the rest frame of the background QSO were used.  We
use 1070 \AA\ as the lower limit to stay well away from the O~VI and
\lyb\ emission lines.  We discuss this at length in \citet{tytler04b}.
Our upper limit of 1200 \AA\ rest is 4000 \kms\ from the QSO, or about
20 Mpc.  This is expected to be well outside the proximity region of
most of our QSOs, and at this distance in no case is the UV radiation
from our background QSOs expected to be more than 20\% of the UV
background intensity.  The UV flux from our median background QSO is only
about 1\% of the UV background at 20 Mpc.

The curves on Figure \ref{abs00ned} show the DA we expect to see
assuming that (1) the foreground QSO radiates isotropically at the
Lyman continuum, (2) the IGM density near the foreground QSO is not
enhanced, and (3) the ionizing flux from the foreground QSO is either
100\%, 10\% or 0\% of the ionizing flux inferred from the QSO
magnitude.  To generate the expected DA curves, we first computed the
expected DA in every pixel of every sightline, and then combined the
background sightlines in exactly the same way as we combined the data,
preserving the $d$ values of the pixels.  We add a uniform background
of DA$_{\rm metal} = 0.02$ to each curve at wavelengths higher than
the \lya\ emission line.  Our expected DA curves are specific to the
QSO luminosities and transverse separations of the QSOs in our data
set.

We see no sign of the expected proximity effect in either the
line-of-sight or transverse directions.  In the transverse direction
we see evidence for enhanced absorption at $-6$ to $0$~Mpc.  There are
fewer and fewer sightlines contributing pixels at increasingly
negative distances than at positive distances, so the quality of our
DA measurements is significantly lower at larger negative distances in
the transverse direction.  In the line-of-sight direction our data is
consistent with no proximity effect.

The expected transverse proximity effect would be more confined and
easier to see if we had smaller redshift errors.  When we calculated
the expected transverse proximity effect that we show in Figure
\ref{abs00ned}, we included random redshift errors with a standard
deviation of 676~\kms . The 676~\kms\ is the dispersion we measured in
the differences between the C~IV and Mg~II redshifts, appropriate
because we took 78 of the 130 foreground QSO redshifts from C~IV.  The
result is that the missing absorption is spread out over many DA bins
instead of being concentrated in the bins near zero.  In Figure
\ref{transversenoz} we show what would be possible if we had zero
error systemic redshifts for all of our foreground QSOs.

\dfigls{los-emission.ps}{transverse-emission.ps} {\label{abs00ned} The
  proximity effect.  The data histogram in the top panel shows DA
  averaged over all foreground QSOs as a function of distance in
  proper Mpc along the sightline, with the origin corresponding to the
  redshift of the foreground QSO.  The data are in bins with width 2.5
  Mpc, or 549~\kms\ at $z=2.2$.  The curves show the expected
  proximity effect in DA for foreground QSO UV fluxes of (top to
  bottom): 0, 10\%, and 100\% of the inferred flux at the Lyman limit.
  The bottom panel shows DA towards the background QSO, with the
  curves showing the expected transverse proximity effect.  We
  calculate the DA we expect at various $d$ (not the 3D distance $r$)
  along each sightline at impact parameter $b$, using the $b$ and $d$
  to calculate the QSO flux.  We then sum these different expected DA
  curves to give the mean expected DA curves that we show, preserving
  distances $d$ along the line-of-sight, in the same way we sum the
  background QSO spectra.  In calculating the expected proximity
  effect we have assumed that the systemic redshifts we calculate from
  our emission lines have random errors of $676$ \kms. This has the
  effect of smearing out the expected proximity effects -- note how
  the zero UV flux line-of-sight proximity effect is not a step
  function in the presence of redshift errors.  }

\dfigl{transverse-0-0.ps} {\label{transversenoz} The expected
  transverse proximity effect in the absence of $z_{\rm em}$
  uncertainties.  As Figure \ref{abs00ned}, but now the expected
  transverse proximity effect has been calculated assuming that there
  are no $z_{\rm em}$ errors.  At $d=0$ our total data set effectively
  has $\omega = 5.2$.}

\subsection{On our error estimates for the binned DA values} 

Each DA bin in Figure \ref{abs00ned} contains contributions from a
large number of sightlines.  For the line-of-sight panel, the typical bin
has contributions from about 100 QSOs.  The number is not
exactly 130 for each DA bin because portions of each spectra are
masked to exclude DLAs, known metal absorption, and regions of bad
data.

The DA bins in the transverse panel of Figure \ref{abs00ned} typically
have contributions from about 40 QSOs at negative velocities and 95
QSOs at positive velocities.  The smaller number of contributions at
negative distances is because our pairs were selected (for a different
project) to have emission redshifts that are close to each other, so
almost every sightline contributes pixels to the transverse DA at zero
and positive distances, while many fewer are able to contribute pixels
to the negative distance bins.

The value plotted for each distance bin is the mean of one DA value from
each of the sightlines with data at the appropriate distance.  The
sightlines are given equal weight.  For each bin, we take the error bar
to be the standard error on the mean: the standard deviation of the
contributing sightlines divided by the square root of the number of
contributing sightlines.  We have also estimated the error for each
bin via bootstrap resampling, and the two methods are consistent with
each other for all bins to better than 10\%.

In Figure \ref{hist2} we show the distribution of DA values measured
over 2.5 Mpc for all of the points that went into the transverse panel
of Figure \ref{abs00ned}.  In Figure \ref{hist1} we show the
distribution of sightline DA values for three of the individual
distance bins.  The distributions do not look like a normal
distribution because the flux probability distribution function (FPDF)
is highly non-normal \citep{tp:kim2007b, tp:tytler2008b}.  We work
with the distribution of the mean fluxes in the 2.5 Mpc bins, so our
distributions are much closer to a normal distribution than they are
to the underlying full-resolution FPDF, as shown for simulated spectra
by \citet[Figs. 10 \& 19]{tp:tytler2008b} and as expected by the central
limit theorem.

The central limit theorem also guarantees that if we take a large
number of samples from the distribution shown in Figure \ref{hist2},
that the mean value with be normally distributed with a standard
deviation given by the standard error on the mean (e.g. our error
bars).  The distributions we observe for individual bins in Figure
\ref{hist1}, as well as the fact that the distribution observed in
Figure \ref{hist2} is not wildly different from a normal distribution,
give us confidence that the errors we have presented in Figure
\ref{abs00ned} are reasonable.
 
\dfig{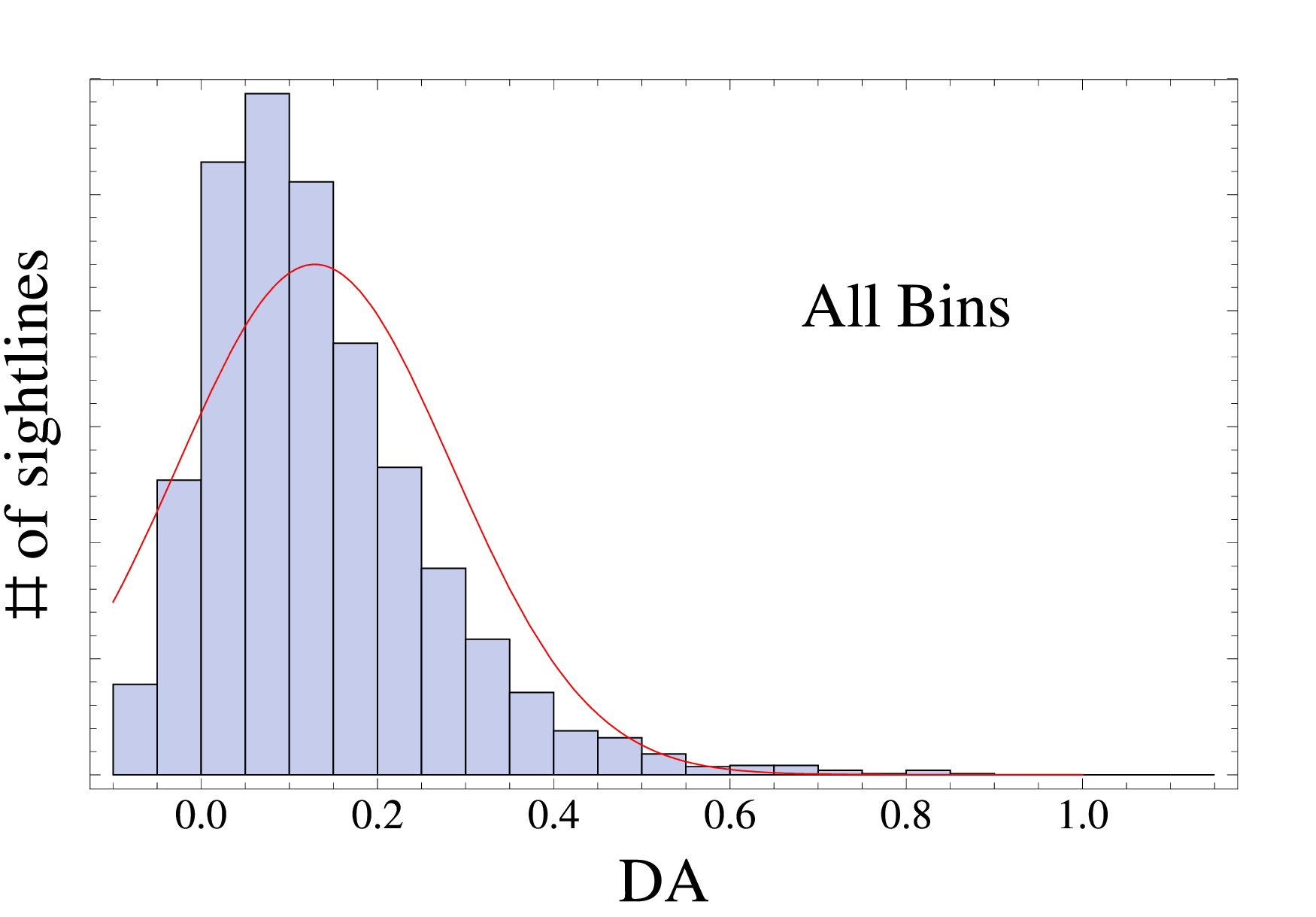}{\label{hist2} The distribution of DA values, each
  given by the mean DA measured over 2.5 Mpc in one sightline, for all
  of the values that are contained in the transverse panel of Figure
  \ref{abs00ned}.  The red curve curve shows a normal distribution
  centred on the mean of these values with the standard deviation
  given by these values.}

\dfig{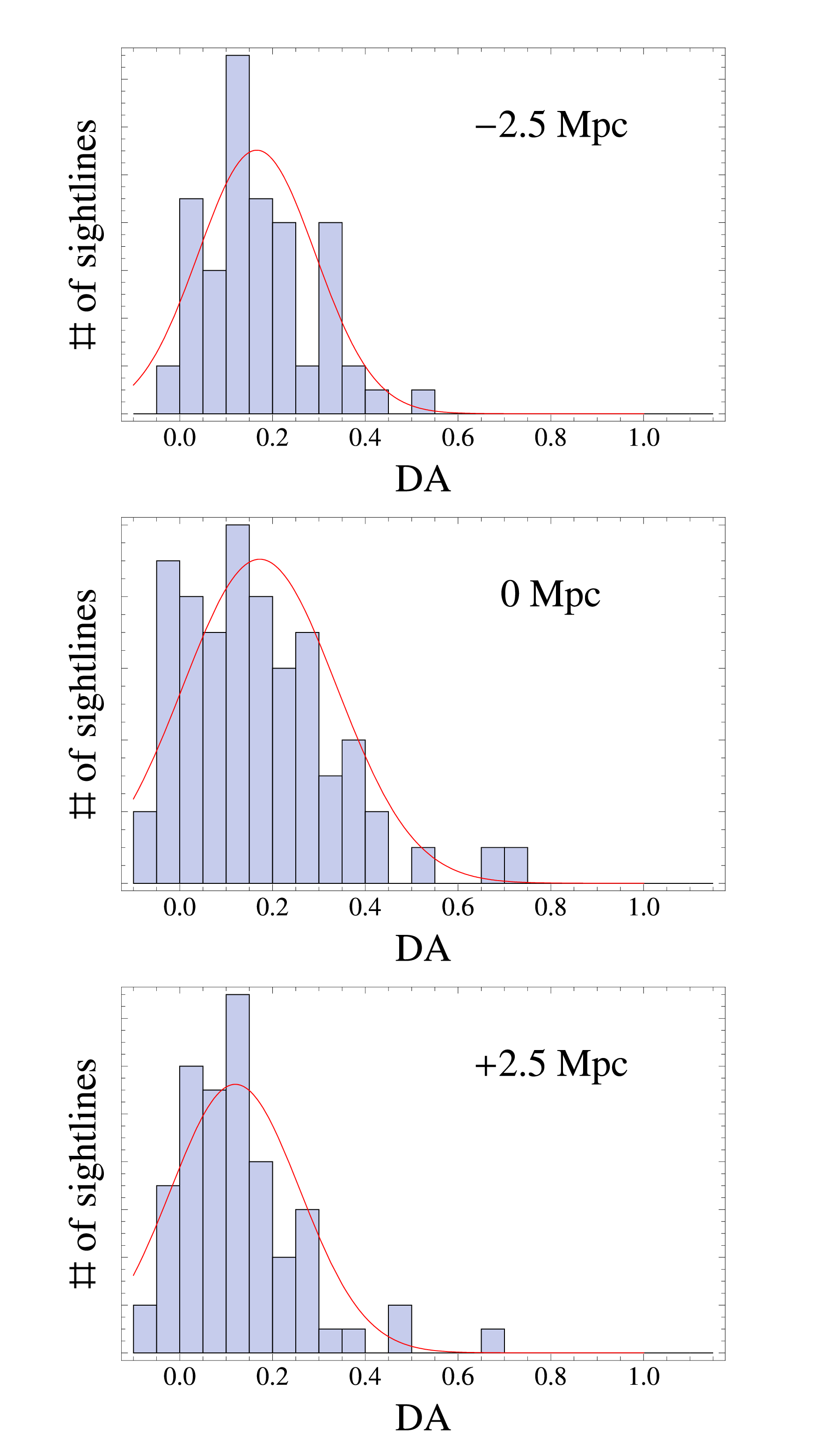}{\label{hist1} As Figure \ref{hist2}, but showing only
  the values contributing to the transverse bins centred at -2.5, 0,
  and 2.5 Mpc.}

We measure the covariance for adjacent 2.5 Mpc bins in our transverse
DA data to be 20\% of the variance in each bin.  The covariance drops
to 5\% two bins out.  This is a bit higher than expected purely from
the large scale structure of the \lyaf\ \citep{kirkman07a}, where we
found that the covariance at 550 \kms\ (about 2.5 Mpc) was 6\% of the
value at 50 \kms\ at $1.0 < z < 1.5$.  The excess covariance in our
sample may be due in part to its higher redshift, but it probably
also reflects errors in our data handling, including continuum fitting
errors and residuals from unmasked LLS and metal line absorption.  The
bin-to-bin covariance is large enough that is should be taken into
account when attempting to determine the significance of any feature
in our DA data.

\section{Line-of-sight Proximity Effect Discussion}

We see in Figure \ref{abs00ned} that we expect a relatively small
line-of-sight proximity effect, because our QSOs are fainter by a
factor of few than those in past work. Our QSOs have a median Lyman
Limit luminosity of 3.3$\times 10^{30}$ ergs s$^{-1}$
Hz$^{-1}$, which is fainter by a factor of six compared to the
\citet{tp:guimaraes2007a} sample.  Four of the 10
\citet{tp:liske2001a} QSOs are more luminous than ours by a factor $>
6.7$, and the others are more luminous by a factor of $\sim 2$.

We still expect to readily detect the proximity effect but we see no
change in the amount of H~I absorption as we approach the foreground
QSOs.  This absence of the line-of-sight proximity effect is
unexpected, but not completely surprising because others have recently
reported reduced effects, and we have a ready qualitative explanation
if the gas density near to the QSOs is enhanced by a factor of a few.

QSOs are expected to form in dense environments.
\citet{tp:serber2006a} find that the galaxy density within 100 kpc of
$z < 0.4$ QSOs is between 1.4 and 3 times the galaxy density around
L$^*$ galaxies, and that the overdensity persists at some level out to
1 Mpc.  The environments around QSOs at $z = 2.2$ may be significantly
different, as clustering will be less developed, and different
types of galaxy may show QSO activity.  IGM calculations indicate
the mean gas density may be enhanced by a factor of a few within about
3 Mpc of a QSO \citep{tp:loeb1995a, tp:faucher-giguere2008a} -- this
is approximately the factor we need to explain our non-detection of
the line-of-sight effect.  The IGM \lya\ opacity will vary
approximately as the square of the gas density -- one factor for the
increased density, and another for the increased neutral
fraction. Hence a factor of 3 increase in density will give about a
factor of 9 more H~I absorption, which will change the expected
proximity effect to approximately that for QSOs with 10\% of their
observed UV luminosities.  The curve in Figure \ref{abs00ned} for this
reduced luminosity is consistent with the data, given the
uncertainties over the redshifts.

\citet{tp:rollinde2005a} also saw less proximity effect than expected
and they deduced that the gas density might be enhanced by ``a factor
of a few'' near to their QSOs.  \citet{tp:guimaraes2007a} saw a
reduced proximity effect towards more luminous QSOs at higher
redshifts and they claim that they need a significant density
enhancement over a much larger region of $\sim 21$ Mpc, more than
expected from simulations.

If the gas density is a factor of a few higher around QSOs and does not
depend on QSO luminosity at a given redshift, then the density
enhancement will have a larger impact on less luminous QSOs because
the distance to which the QSO flux dominates the UVB flux is then
smaller than for more luminous QSOs.  This might explain why earlier
papers \citep{carswell82,bajtlik88,tp:scott2000a} readily saw the
line-of-sight proximity effect around QSOs that were more luminous
than ours.  
However, the different results might instead come from the different
methods. Our methods and those of \citet{tp:rollinde2005a}
and  \citet{tp:guimaraes2007a}, who also claimed enhanced density near to QSOs,
are based on flux, while the early papers that saw the expected
line-of-sight proximity effect used line counting.
To our knowledge, no one has attempted
a proximity effect analysis on the same data set using both line
counting and continuous optical depth methods, so it is possible that
the two methods give systematically different results.

\subsection{Other possible explanations for why we do not see a
                    line-of-sight proximity effect}
\label{nolosexp}

While we are happy to entertain the idea that enhanced density
explains why we do not see the line-of-sight proximity effect that we
expect, we have not directly shown that this is the case.  It remains
surprising that the enhanced ionization -- density cancellation is
perfect within our measurement errors.  Here we explore other possible
explanations for what we see.

In addition to enhanced density near our foreground QSOs, it may be
possible to explain our lack of an observed proximity effect in other
ways.  In particular, Figure \ref{abs00ned} shows that the
line-of-sight data is roughly compatible with $\omega$ values 10 times
smaller than we expect.  This could be achieved, for example, if the
UV background is a factor of 10 higher than we expect.  However, our
work on the UVB from \lyaf\ absorption \citet{tytler04b, jena05a,
  tp:tytler2008b}, the \citet{bolton05a} results using similar
methods, and the \citet{haardt01a} result derived by counting UV
sources all suggest that $\Gamma_{\rm UVB}$ is less than a factor of
two higher than the value of $\Gamma_{\rm UVB} = 1.3 \times 10^{-12}$
s$^{-1}$ that we have adopted.  We believe that the 50\% error claimed
by Bolton is more reasonable than a factor of two, and hence we think
that it is unlikely that our result will be explained by a higher than
expected UVB.

The flux enhancement near the foreground QSOs, $\omega$, could be
lower than we expect if these QSOs were less luminous in the recent
past than they are today.  At a distance from a QSO where the QSO UV
radiation is twice that of the UVB, we expect it to take 10 kyr ($(2
\Gamma_{\rm UVB})^{-1}$) for the ionization of the gas to respond to
increased UV flux from the QSO, so $\omega$ can be different than we
expect if QSOs are highly variable on timescales short compared to 10
kyr.  We are presumably more likely to discover and observe bright
QSOs, so if QSOs are varying on short timescales the sense of this
effect may on average lower $\omega$.

On time-scales of days to years, QSOs are more variable at smaller UV
wavelengths and at lower luminosities \citep{tp:vandenberk2004a}.
Over tens of years in the rest frame, the rest frame UV flux varies by
$> 1.5$ magnitudes for 50\% of QSOs and by $>3$ mag for 9\% of QSOs
\citep{tp:heckman1976a}. \citet{tp:devries2003a,tp:devries2005a} find
that QSOs undergo bursts of 2-magnitudes on periods of years, with
larger variations on longer time scales and for less luminous QSOs,
all roughly consistent with accretion disk instabilities.  The
structure function describing the variability of all QSOs rises
monotonically at a constant rate out to 40 years in the QSO rest frame
with no turnover \citep{tp:devries2005a, tp:devries2006a} -- the
preferred time scale for QSO variation is at least this long.  If the
same random walk has continued, then many QSOs could be factors of ten
times less luminous 10~kyr ago.  But \citet{tp:martini2003a} estimate
that QSO UV luminous episodes typically last $>20,000$ years, on the
assumption that a given QSO seen on the POSS-I plates was either on or
off (absolute magnitude fainter than -23) at the epoch of the SDSS
observation.

Our adopted redshifts may differ systematically from the QSO systemic
redshifts. To give the full expected proximity effect we would need to
increase the QSO redshifts by 800~\kms , moving the origin on the
figures to the left by 1.5 data bins or 3.5~Mpc.  Given that our
redshifts come either directly, or indirectly from Mg~II, this seems
an incredibly large error, which would imply that the C~IV lines in
our QSOs have typical blueshifts of 1550~\kms\ and not the 753~\kms\
that we assume.

Another alternative explanation is that part of the line-of-sight proximity
effect is masked by extra H~I absorption that is not from the IGM.  It
is well known that there is an excess of absorption systems that show
C~IV lines with redshifts similar to QSO emission redshifts.
\citet[Fig. 7a]{tp:tytler2008a} showed these excess systems for a
super-set of the QSOs that we use here. \citet{tp:hennawi2006a} show
there is also a strong excess of LLS and DLAs with redshifts similar
to the QSOs. In addition \citet{tp:wild2008a} find that $>$ 40\% of
C~IV absorbers within 3000 \kms\ of a QSO are directly associated with
the QSO itself and do not arise in the IGM.  These systems will nearly
all have strong \lya\ absorption lines that will tend to hide the
line-of-sight proximity effect.

In general, the extra absorption near the QSO redshift is a specific
example of the idea that the gas density is higher near to the QSOs.
In early work on the proximity effect
\citep{carswell82,tp:tytler1987a,bajtlik88}, all \lya\ lines that had
associated metal lines were excluded, hence removing this non-IGM
``contamination'' at all redshifts.  This was not done in later work
\citep{tp:scott2000a, scott02, tp:dallaglio2008a}.  \citet[\S\
6]{scott02} found that removing ``associated absorbers, damped \lya\ 
absorbers, and blazars'' from their low redshift proximity effect
analysis decreased the UVB needed to explain the proximity effect by a
factor of two.  This factor of two reduction seems desirable
\citep[Fig. 1]{tp:faucher-giguere2008a}, because the UVB from the
proximity effect then matches that inferred by matching the observed
DA to large hydrodynamic simulations \citep{bolton05a,
  tp:tytler2008b}.

For this paper we have attempted to remove only part of the \lya\
absorption associated with metal systems, that where the \lya\ lines
are DLAs or other prominent \lya\ lines. We have not removed the \lya\
lines of other metal systems, and we do not know whether these could
significantly or totally cancel out the line-of-sight proximity
effect.

We could also speculate that our QSOs might show more than the typical
amount of extra metal systems with redshifts similar to the QSO
redshifts, perhaps because they are lower luminosity QSOs. This is
hinted because X-ray absorption is more common in lower luminosity
AGN.

\section{Transverse Proximity Effect Discussion}

We now describe the transverse proximity data in Figure
\ref{abs00ned}, and we discuss the issues that carry across from our
interpretation of the line-of-sight proximity effect. We end with new
factors that are specific to the transverse proximity effect.

The first point to make is that the spectra that we sum for the
transverse plot are the different than the ones we use for the
line-of-sight.  In the transverse direction we use the spectrum of the
background QSO, while for the line-of-sight we use the spectrum of the
foreground QSO.  Hence the noise characteristics are similar, but not
identical.  As we previously explained, we lose spectra as we move
farther behind the foreground QSOs on the transverse plot, which
explains why the errors are nearly constant to the right but increase
going to the left of zero.

The second point is that we see no change in the amount of absorption
as we approach the QSOs from the Earth side, from the right.  This
seems reasonable because we also did not see any change in the
absorption in the line-of-sight to these same QSOs. The same
explanation that we gave for the lack of the line-of-sight proximity
effect may apply to the lack of change in the transverse absorption,
because we are probing similar distances with both the foreground and
the background QSO light.  Hence we propose that we do not see either
the line-of-sight proximity, or the transverse proximity on the near
side of the foreground QSO, because the enhanced from ionization the
UV flux from the foreground QSO is cancelled by higher gas density near to
those QSOs.

We discussed other possible explanations for the lack of the
line-of-sight proximity effect. While arbitrarily large systematic
errors in the redshifts of the foreground QSOs might enhance, diminish
or remove the line-of-sight proximity effect, redshift errors have
much less effect on the transverse effect, because we can now also
see absorption from behind the foreground QSO.  Systematic errors in
the foreground QSO emission redshifts will again move the zero point
to the left or right on the plot, but this has little effect because
the absorption we observe and expect does not change significantly
when we apply realistic shifts in the zero point.

Random redshift errors do not change the total amount of absorption,
but they do re-distribute that absorption into more pixels. In Figure
\ref{transversenoz} we re-calculate the expected transverse proximity
effect assuming that we have no errors in the QSO redshifts. We see a
deeper and narrower expected drop in H~I absorption.

Extra \lya\ absorption for the excess of metal line systems with
redshifts close to the redshift of the background QSO will have little
impact on the transverse proximity effect, except when the two QSOs
have similar redshifts.  The extra absorption near to the foreground
QSO is one manifestation of the enhanced density that we believe is
important.  Associated absorbers ejected by the foreground QSOs are
not expected to reach the sightline to the background QSO.

The third point about the transverse plot is that we see extra
absorption starting at the foreground QSO position and extending about
6 Mpc behind the foreground QSO. We regard this as significant for two
reasons. First, we see a 2 -- 3 $\sigma $ excess over 3 pixels,
extending from $-6.25$ to $+1.25$~Mpc.  Second, we had earlier decided
that the lack of the expected line-of-sight proximity effect was
significant, and that the lack of the expected transverse proximity
effect on the front side of the foreground QSOs was also
significant. These two lacks involve approximately the same deviations
from the data as does the excess absorption behind the QSOs.  Hence we
should also regard the excess absorption as significant.  This
argument relates to the Bayesian preference that we not change our
prior evaluation, of what would constitute a significant result, after
we see the data.  Rather we should hold a consistent set of beliefs
about probabilities before and after we obtain the data, striving for
diachronic probabilistic coherence.  While we have a significant
detection of excess absorption, we are less sure of the precise
location of the excess because random and especially systematic
redshift errors can move the apparent location of the absorption. At
least some of the excess is behind the foreground QSOs.

The asymmetry between the amount of absorption in front of and behind
the foreground QSOs could only be seen in the transverse analysis,
because the line-of-sight analysis is only sensitive to absorption in
front. We also require a large sample of close pairs of QSOs to see
this effect, with emission redshifts at least as good as we have.  It
is clear from Figure \ref{abs00ned} that it would be hard to see this
asymmetry in a much smaller sample.

\section{Implications  of  anisotropic absorption}

The transverse sightlines in Figure \ref{abs00ned} suggest that there
is more absorption behind the foreground QSOs than there is in front.
If this result is correct it may have significant implications, some
of which we now discuss.

The amount of excess absorption behind the QSOs is numerically similar
in size to the lack of absorption that we had expected. This
immediately suggests that the excess absorption is coming from an
enhancement of the gas density that is the same size as the
enhancement that we already invoked on the near side of the QSO. We
expect the density distribution to be isotropic about the QSOs, when
we average over many QSOs. We can then explain the enhanced absorption
using the same density enhancement, but with no UV radiation
from the foreground QSO reaching the gas behind the foreground QSOs
before the absorption occurred.  We can not say precisely how much the
flux behind the foreground QSOs needs to be suppressed, but we can see
from Figure \ref{abs00ned} that if $\omega$ is down by about a factor
of 10 behind the QSO, and we have a symmetric density enhancement
centred on the QSOs, then we would expect to see something similar to
our observed data.

There are two commonly discussed ways of limiting the amount of UV
flux seen by the gas behind the foreground QSOs. First, the QSO
emission might be anisotropic.  Second, the QSO may have a short
episodic lifetime.  We will discuss both possibilities.

\subsection{Anisotropic UV emission?}

Common AGN unification models frequently contain an obscuring torus
surrounding a central continuum source and broad emission-line gas
(e.g.  \citep{tp:barthel89a, tp:antonucci93a}).  In this scenario, QSO
UV emission is expected to be highly anisotropic, with the UV emission
strongest along the poles of the system when the obscuring torus
defines the equator.  But while an obscuring torus could explain a
general lack of ionizing photons in the transverse direction, it does
not explain why there may be fewer ionizing photons behind the QSO
than in front, because the UV radiation should escape
equally from both sides of the torus.

We can break the axial symmetry in the UV emission if the obscuration
around the QSOs covers most sight lines, leaving only a few holes
unobscured, including the hole sending UV in our direction, a
modification of the cloudy torus model of \citet{tp:nenkova08a}. A
single hole of diameter of order 60 -- 120 degrees seen from the QSO
might explain our data.  Here 60 degrees is the minimum to illuminate
enough of the volume in front of the QSO, while larger than 120
degrees leads to too much flux behind the QSOs since the line-of-sight
to us is often far from the centre of the hole. This model is
effectively similar to a hypothetical accretion disk that emits UV
from one side but not the other. Models with several smaller
unobscured holes are not favored because they do not give much less
flux behind the QSOs.

We might explain the excess absorption behind the QSOs if the UV flux
behind the average QSO is of order 10\% or less of the flux we
see. A single unobscured hole of diameter 60 -- 120 degrees covers a
fraction of 0.25 -- 0.43 of the sky seen from a QSO.  We expect this
fraction is related to the fraction of all QSOs that are type 1 rather
than type 2. For Seyfert galaxies, \citet{tp:schmitt2001} estimate 0.3
while \citet{tp:hao2005} find 0.5. This fraction rises with luminosity
reaching of order 0.8 for QSOs \citep{tp:maiolino2007,tp:barger2005}. We might
reconcile this high fraction of unobscured QSOs with our need for more
obscuration because we need block only the Lyman continuum flux, and
not the entire UV and optical.  The obscuration we need does not
necessarily lead to high mid-infrared to optical flux ratios used to
calculate the fraction of 0.8.

\subsection{Short episodic lifetimes?}
\label{eplifesec}

A second option is that the QSOs have not been emitting for long
enough to have illuminated the volume probed behind. In this case the
asymmetry is caused by the extra time for the UV radiation from the
foreground QSO to reach the gas behind, and to do this before the light
from the background QSO passes through that gas.

We can account for the excess absorption behind the foreground QSOs if
they have had their current UV luminosities for approximately 1 Myr,
and prior to then, for $> 40$~Myr, they were a factor of $\sim 10$
less luminous. We work in the QSO frame, so that ``today'' refers to
the time in the QSO frame when the light that we see left the QSO;
hence we ignore the time for light to travel from the QSO to us, and
we can ignore the $(1+z)$ time dilation would apply if we were to
shift to our frame.  Our closest sightlines are separated by $\sim
0.1$ Mpc, corresponding to a light propagation time of $\sim 0.32$
Myr, while our median sightline is separated by 1.25 Mpc or 4.1
Myr. The typical foreground QSO must have emitted the flux that we deduce
for at least 4.1 Myr if that flux is to reach the closest approach of
the line-of-sight from the background QSO.  Longer is needed to
illuminate the parts of the background QSO line-of-sight that are
behind (at higher redshift than) the foreground QSO.  The surface that
is illuminated by radiation that left the foreground QSO $t$ Myr ago
is a paraboloid with the QSO at the focus and the vertex $t/2$ million
light-years behind the QSO, as shown in
\citet[Fig. 3]{tp:adelberger2004a}, \citet[Fig. 1]{tp:visbal2008a},
and \citet[Fig. 24]{tp:tytler2008a}).  We demonstrate this in Figure
\ref{figdelay}.

We can obtain approximate limits on the QSO episodic lifetime from the
distribution of absorption around the foreground QSOs.  If the QSOs
had at least their current luminosity for more than 10 Myr, then
regions that are approximately 1.2 Mpc behind the foreground QSO (and
at the median sky separation) would have experienced the QSO flux, and
we might expect that we would not see excess absorption in the bin
centred on zero.  Hence we can deduce, because of the excess
absorption in the zero bin, that the typical QSO episodic lifetime is
$< 10$ Myr.

On the other extreme, we note that the bins in front of the QSO do not
show enhanced absorption.  If we assume that the IGM density enhancement is
symmetric in front of and behind the foreground QSOs, the lack of
extra absorption in the bins in front of the foreground QSO can be
taken as evidence that they have been illuminated by the QSO.  For our
median separation, it will take $\sim 0.3$ Myr for the QSO to
illuminate the bin sampling 1.25 -- 3.75 Mpc in front of the QSO.
Hence we can deduce that the typical QSO episodic lifetime is $> 0.3$
Myr.

Taken together, the transverse absorption in front of and behind the
foreground QSOs suggest a QSO episodic lifetime $0.3 < t_e < 10$ Myr,
or $ t_e \sim 1$ Myr.  Again, a given QSO might have several or many
epochs with high UV luminosity.  Hence the 1 Myr episodic lifetime
refers to the time since the start of the latest QSO outburst and not
to the total QSO lifetime.  

This model also gives the approximate minimum time the QSOs should
have been in their low UV luminosity ``off states'' prior to the
current UV bright episode. To avoid illuminating the bin furthest
behind the QSOs with excess absorption at $d= -6.25$ to $-3.75$~Mpc at
the typical $b=1.25$~Mpc, we need the QSOs in the off state for $>
40$~Myr.  If we knew the distribution of density near the QSOs, and we
had more accurate redshifts, we could make a more accurate estimate of
how long the QSOs have had their current UV luminosities, and the
minimum length of the off state.

Our conclusion that we are seeing evidence for $\sim 1$ Myr QSO
episodic lifetimes hinges on our observation of increased absorption
behind the QSO but not in front of it.  In an earlier analysis of the
transverse proximity effect, \citet{tp:croft2004a} found enhanced
absorption on both sides of the QSO.  He had few close sightlines and
convolved his observed absorption with a 7~Mpc Gaussian filter, but it
is clear that the excess absorption is centred near to the redshift
of the foreground QSO.  We can not explain the discrepancy between our
results and the \citet{tp:croft2004a} results, but the differences
between our assumed systemic redshifts and the systemic redshifts used
by SDSS may be part of the difference.  

We note that if we were to increase our foreground redshift by 550
\kms\ to centre the excess absorption at 0 Mpc, we would observe a
significant line-of-sight proximity effect.  But we would then have
the problem that on average, our C IV emission lines would be at the
systemic redshift of the QSO, and we discussed in \S \ref{redshifts}
this is inconsistent with multiple observations. We also think this
unlikely because \citet{tp:tytler2008b} see excess C~IV absorbers at
velocities $\sim 300 \pm 150$~\kms\ in front of a superset of the QSOs
we study here.  If we move all the QSOs back 550~\kms\ these excess
C~IV absorbers would be at $\sim 850$~\kms\ in front of the QSOs,
which would be hard to explain.  Rather, the C~IV absorption suggest
that we might move the QSO redshifts in the other direction,
decreasing them by $\sim 300$~\kms\ to approximately centre the C~IV
on the QSOs. This would move the origin in Figure 4 one half bin to
the right, putting the extra H I absorption entirely behind the QSOs.

\subsection{Are short episodic lifetimes plausible?}

It has long been speculated that QSOs may exhibit highly intermittent
activity.  \citet{tp:shields1978a} showed that the storage and release
of gas in an accretion disk would produce just such activity.  Disk
instabilities are likely to produce variation on many time scales, from
$10^{1}$ -- $10^6$ yrs \citep{tp:wallinder1992a}.  But it
is also the case that the characteristic e-folding time for a black
hole to increase its mass by accretion is 45 Myr when its
luminosity is the Eddington value and the radiative efficiency is 0.1
(see review by \citet{tp:martini2004a}).  If such accretion is the
dominant mode of black hole growth, then we expect a given high mass
black hole to be UV luminous for of order $10^8$ yr in total, but we
do not know whether this manifests as one long episode or many
shorter bursts.

\citet{tp:goodman2003a} finds that there are no obvious ways to have
accretion disks around giant black holes that are stable against
fragmentation at large radii. Consequently, he suggests that disks do
not exist much beyond 0.01 pc or 1000 Schwartzschild radii.  Such
small disks would be depleted onto the black hole in $<1$ Myr,
implying the typical UV luminosities 1 Myr ago could have been at
least an order of magnitude less, as we require to explain the excess
\lya\ absorption behind the QSOs.

The excess absorption seen behind the QSOs might be a detection of
instabilities in accretion disks on Myr time-scales
\citep{tp:czerny2006a}.  \citet{tp:janiuk2004a} discuss how the
thermal-viscous instability in standard $\alpha -$accretion disks can
lead to outbursts of $10^4$~yr duration for a $10^8$ \msun\ black
hole.  \citet{tp:mchardy2006} support the old idea
\citep{tp:shakura1976} that AGN are scaled up Galactic stellar
mass black hole systems because they can predict the time scale of
the break in the power spectrum density of the X-ray fluxes from the
black hole mass and luminosity or accretion rate (see also
\citet{tp:fender2007}).  \citet{tp:done2005a} discuss how the
transitions between the different states of accretion luminosity seen
in Galactic stellar mass black holes may manifest in $10^9$ \msun\
QSOs as transition into low luminosity states on time-scales of
0.3~Myr.

If we combine these arguments with the large amplitudes of the
observed short-term variability, the lack of a turn-over in the
structure function out to tens of years, the increase in variability as
smaller UV wavelengths (all discussed in \S \ref{nolosexp}), we find
it reasonable to postulate that our sample of QSOs have episodic
life-times $\sim 1$~Myr in the QSO rest frame.  Prior to the current
episode, for perhaps tens of Myrs, their Lyman continuum luminosity
was at least a factor of ten less than today.

\dfig{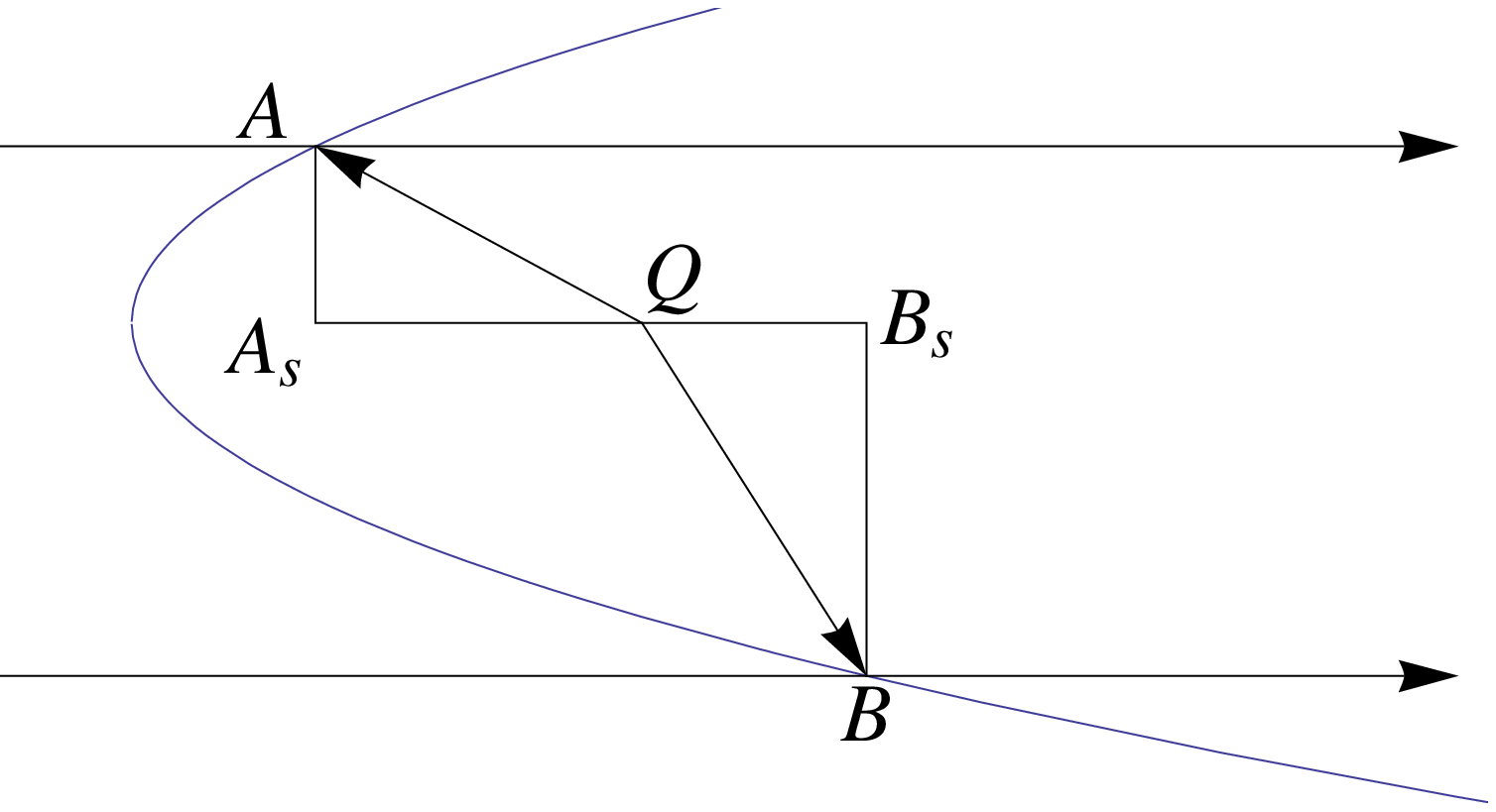}{\label{figdelay}The regions of space around a QSO
  that will show enhanced ionization in the spectra of background
  objects, if the QSO has some fixed episodic lifetime, and the medium
  is at all times highly ionized.  Absorbers to the right of the blue
  parabola will have seen enhanced ionization from the QSO before the
  light from the background object arrives. Light from the background
  objects (horizontal lines) will arrive at absorbers to the left of
  the parabola before the UV flux from the QSO arrives. In the frame
  of the QSO, light from a background object will arrive at point $A$
  a time in the past equal to the light travel time from $A_s$ to $Q$,
  which we denote as $t(A_sQ)$. We take the light travel time as
  positive. So point $A$, on the boundary of the illuminated volume,
  can be found by requiring that $t(AQ) + t(A_sQ) = T$, where $T$ is
  the episodic lifetime of the QSO.  The same argument applies for
  $B$, but because light from a background object arrives at point $B$
  at a time in the future with respect to the QSO frame, we have
  $t(BQ) - t(QB_S) = T$.  Taking the set of points similar to $A$ and
  $B$ at all impact parameters gives the blue parabola.  Light emitted
  from the foreground QSO $T$ years ago, arrives on the parabola at
  the same time as the light from the background QSO that we see.}

\section*{Acknowledgments}

We thank the anonymous referee for many very helpful suggestions and
criticisms.  We thank Arlin Crotts for providing us with some of the
spectra used in this paper, which he obtained from NOAO facilities,
operated by AURA for the NSF. Former UCSD students John O'Meara and
Nao Suzuki helped to obtain some of the spectra used in this paper. We
thank Mike Fall, Jeremy Goodman, Matt Malkan, Jason Prochaska, Don
Schneider, Greg Shields, Chuck Steidel and Gordon Richards for
important discussions. The bulk of the data were obtained using the
Kast spectrograph on the Lick Observatory 3m-Shane telescope, the LRIS
spectrograph on the Keck-I telescope, and the SDSS archive.  The
W.M. Keck Observatory is operated as a scientific partnership among
the California Institute of Technology, the University of California
and the National Aeronautics and Space Administration and was made
possible by the generous financial support of the W.M. Keck
Foundation. We are exceedingly grateful for the help we receive from
the staff at both observatories. We recognise and acknowledge the very
significant cultural role and reverence that the summit of Mauna Kea
has always had within the indigenous Hawaiian community.  We are
extremely grateful to have the opportunity to conduct observations
from this mountain.  This research has made use of the NASA/IPAC
Extragalactic Database (NED) which is operated by the Jet Propulsion
Laboratory, California Institute of Technology, under contract with
the National Aeronautics and Space Administration.  Funding for the
creation and distribution of the SDSS Archive has been provided by the
Alfred P. Sloan Foundation, the Participating Institutions, the
National Aeronautics and Space Administration, the National Science
Foundation, the U.S. Department of Energy, the Japanese
Monbukagakusho, and the Max Planck Society. The SDSS Web site is
http://www.sdss.org/.  This work was funded in part by NSF grants
AST-0098731, 0507717 and 0808168 and by NASA grant NAG5-13113.

\bibliographystyle{mn2e}
\bibliography{archive,tp}

\end{document}